\documentclass[twocolumn]{aastex63}

\newcommand{\reviewchange}[1]{\textcolor{red}}

\usepackage{graphicx,xcolor}
\usepackage{lipsum}
\usepackage{gensymb}
\usepackage{placeins}
\usepackage{float}
\usepackage{hyperref}
\usepackage{booktabs,multirow,tabularx} 
\usepackage{amsmath}  
\usepackage{graphicx} 
\usepackage{tabulary,capt-of}
\usepackage{booktabs}
\usepackage{rotating}
\usepackage{booktabs} 
\usepackage{tablefootnote}
\makeatletter
\newcommand\footnoteref[1]{\protected@xdef\@thefnmark{\ref{#1}}\@footnotemark}
\makeatother
\usepackage{float}
\usepackage{lineno}

\shorttitle{\textit{NuSTAR} view of Perseus}
\shortauthors{Creech, Wik et al.}

\begin{document}

\title{The \textit{NuSTAR} View of Perseus: the ICM and a Peculiar Hard Excess}


\author[0000-0002-9041-7437]{Samantha Creech}
\affiliation{Department of Physics \& Astronomy, The University of Utah, 115 South 1400 East, Salt Lake City, UT 84112, USA}
\affiliation{Astrophysics Science Division, SURA/GSFC/CRESST II, Greenbelt, MD 20771}

\author[0000-0001-9110-2245]{Daniel R. Wik}
\affiliation{Department of Physics \& Astronomy, The University of Utah, 115 South 1400 East, Salt Lake City, UT 84112, USA}

\author{Steven Rossland}
\affiliation{Department of Physics \& Astronomy, The University of Utah, 115 South 1400 East, Salt Lake City, UT 84112, USA}

\author[0000-0002-3132-8776]{Ay\c{s}eg\"{u}l T\"{u}mer}
\affiliation{Kavli Institute for Astrophysics and Space Research, Massachusetts Institute of Technology, 77 Massachusetts Avenue, Cambridge, MA 02139, USA}

\author[0000-0002-5267-2867]{Ka-Wah Wong} \affiliation{Department of Physics, SUNY Brockport, Brockport, NY 14420, USA}

\author[0000-0002-6413-4142]{Stephen A. Walker} \affiliation{Department of Physics and Astronomy, University of Alabama in Huntsville, Huntsville, AL 35899, USA}

\begin{abstract}

As the brightest galaxy cluster in the X-ray sky, Perseus is an excellent target for studying the Intracluster Medium (ICM), but until recently, its active galactic nucleus (AGN) made studies of the diffuse emission near its center nearly impossible to accomplish with \textit{NuSTAR} due to the extended wings of \textit{NuSTAR}'s PSF.
The development of a new open source software package---nucrossarf---now allows the contribution from point and diffuse sources to be modeled so that scattered light from the AGN can be accounted for.
Using this technique, we present an analysis of diffuse hard X-ray (3-25~keV) emission from the ICM using three archival \textit{NuSTAR} observations of the Perseus cluster.
We find a $\sim 10\%$ excess of emission beyond 20~keV not describable by purely thermal models.
By performing similar analyses of AGN in archival observations, we have characterized the systematic uncertainty of the modeled AGN contribution to be 3.4\%.
However, in order to explain the excess, the total scattered AGN emission would have to be 39\% stronger than we have measured.
We test physical explanations for the excess, such as diffuse inverse Compton emission potentially originating from the radio mini-halo, but we determine that none of the models are compelling.
An upper limit on inverse Compton flux ($\leq1.5\times10^{-11}$~erg~s$^{-1}$~cm$^{-2}$) and a corresponding lower limit on global magnetic field strength ($\geq 0.35~\mu G$) is derived.
We discuss the potential origin and implications of the excess and present our characterization of the \texttt{nucrossarf} systematic uncertainty, which should be useful for future work. 

\end{abstract}


\section{INTRODUCTION}\label{sec:intro}

Galaxy clusters are the largest virialized structures in the universe, and their galaxies are immersed in a hot, ionized plasma---the intracluster medium (ICM)---that emits at X-ray energies via thermal bremsstrahlung.
The brightest galaxy cluster in the X-ray sky is Perseus (Figure \ref{fig:glob}) at z~=~0.018 \citep{zPers}, and this
proximity has allowed the investigation of the ICM and its structure in great depth \citep[e.g.,][]{Pers_Chand_Sanders07,Fabian06,Fabian11,Fabian15}. One of the most important findings of the previous studies is the cluster's prominent cool core and bubbles blown by past Active Galactic Nuclei (AGN) activity.
Cool cores are typically formed in dynamically relaxed clusters (i.e., clusters that have not recently undergone a major disruption to their ICM, such as a merger event) as the dense, central plasma cools radiatively more quickly than the rest of the gas in the halo.

Radio observations reveal diffuse and faint synchrotron emission associated with the cool core-region of some relaxed galaxy clusters \citep[e.g.,][]{GM17,minihalo-Pheonix} known as \textit{mini-halos}. 
Mini-halos are treated separately from the large-scale, merger induced extended radio sources---radio halos and radio relics---that are not directly associated with cluster galaxies \citep[e.g.,][]{minihalo_turbulence_model_Perseus_Gitti02}.
Unlike radio halos and relics,
these mini-halos are confined to cluster cores, often bounded by cold fronts, in clusters without other obvious merger signatures.

It is thought that the relativistic electrons forming radio halos are accelerated by the turbulence that develops in the ICM during major merger events \citep[e.g.,][]{Halo_Tribble93,B_coma_Brunetti01}.
Similarly, the particles in radio relics \citep{Relic_class81} are thought to get accelerated by shock waves \citep[e.g.,][]{Relic_formation02}.
Unlike radio halos, these relics are elongated, irregularly shaped structures that appear in the periphery of the ICM.

Barring re-heating processes, radio relics and giant radio halos should not be observable after $\sim 10^8$~yr \citep{minihalo_turbulence_model_Perseus_Gitti02}.
However, some relaxed clusters with cool cores host a mini-halo that extends up to $\sim$500 kpc beyond the radio galaxy at the cluster's center. 
These objects are distinct from giant radio halos; they only appear in relaxed clusters with cool cores.
The first of these objects was discovered in the Perseus cluster \citep{Burns92_minihalo_disc}, and Perseus has since been a prime specimen for studying the physics of mini-halos.

The same pool of relativistic electrons responsible for synchrotron emission in radio structures is expected to uplift cosmic microwave background (CMB) photons to X-ray energies via inverse Compton (IC) scattering.
The strength of IC emission depends on the number density of the relativistic electron populations. 
In the case of radio mini-halos, the synchrotron emission is notoriously faint, which suggests that the electron density is low, especially given the expectation that the magnetic field is strongest in cluster centers.

The synchrotron flux from a population of relativistic electrons is related to the magnetic field strength of the environment, but attempts to estimate the magnetic field using only synchrotron emission rely on the assumptions of equipartition. 
However, if a detectable IC signal is present, one can derive the average magnetic field strength from the ratio of the IC and synchrotron luminosity without evoking these assumptions:
    \begin{equation} \label{eq:sync_IC}
    \frac{L_{\text{IC}}}{L_{\text{sync}}} \propto \frac{u_\text{phot}}{u_\text{B}}\, ,
    \end{equation}
where $u_\text{B}$ is the magnetic field energy density and $u_\text{phot}$ is the energy density of the CMB, which is the seed photon field for IC scattering. 
Unlike equipartition or Faraday rotation methods [see \cite{B_review_04} for an overview of these topics], using IC to measure magnetic field strength is unambiguous and does not rely on assumptions (as in equipartition) or an incomplete knowledge of gas structure (as in Faraday rotation).
However, previous searches for IC---which have largely focused on merging and disturbed clusters with bright radio structures---have been inconclusive, so only lower limits to magnetic field strength have been found so far \citep[e.g.,][]{FirstComa_IC_04,DanComa,DanBullet,RB21,RB23,Mirakhor22,Tumer23}.

Detecting IC is tricky because the thermal ICM mostly dominates the X-ray spectra until $\gtrsim$15--20~keV. 
\textit{NuSTAR} is the only existing imaging X-ray telescope capable of detecting a signal at these hard energies; the sensitive region for diffuse sources is $\sim$3--30~keV. 
Unfortunately, the instrumental background at $\gtrsim$20~keV has rendered many past IC searches inconclusive.

Since mini-halos are much fainter than large-scale halos, the relativistic electron density is expected to be much lower than that of large-scale halos. 
Therefore, mini halos are not of great interest for IC searches in ICM.
However, the Perseus cluster's mini-halo \citep[12.64 Jy at 230--470 MHz;][]{GM17} is brighter by $\sim 3-4$ orders of magnitude when compared to average mini-halos \citep{minihalo_fluxes_Giacintucci14}. 
Thus, Perseus is the best candidate for searching for IC emission from a mini-halo. 

This study reports on \textit{NuSTAR} observations of the core of the Perseus Galaxy Cluster, which is at redshift $z=0.018$ \citep{zPers} and has an X-ray luminosity of $2.19 \times 10^{45}$~ergs~s$^{-1}$ \citep[2-10~keV,][]{Mushotzky78_Pers_Xlumin}, making it the brightest X-ray cluster in the sky. 
The virial mass and radius derived from X-ray measurements are $M_{\odot} = 6.65 \times 10^{14}~M_{\odot}$ and $r_{200} = 1.79$~Mpc, respectively \citep{Simionescu11_Pers_Pars}. 
Perseus is a cool core cluster with a central temperature of $\sim$3~keV that rises to $\sim$5.5~keV at 120~kpc \citep{Schmidt20_Pers_Chandra_TempMap} with a central X-ray emitting AGN and a radio mini-halo.

Throughout this paper, we assume a $\Lambda$CDM cosmology with {\it H$_{0}$} = 70 km s$^{-1}$ Mpc$^{-1}$, $\Omega_{M}$ = 0.27, and $\Omega_{\Lambda}$ = 0.73. Under these assumptions, the angular scale at the redshift of the Perseus Cluster is $\sim$0.37~kpc/$\arcsec$. All uncertainties are quoted at the 3$\sigma$ confidence levels unless otherwise stated.

In this study, we report on our search for an IC signal in the ICM of the Perseus Galaxy Cluster.
In Section \ref{sec:DR}, we describe observations and data reduction processes. 
In Section \ref{sec:analysis}, we discuss the spectral analysis, our treatment of systematic uncertainties, and the characteristics of the hard excess. 
In Sections \ref{sec:disc} and \ref{sec:conc}, we discuss our results and summarize our findings.
Lastly, derivation of the systematic uncertainty associated with scattered emission from the AGN is described in an Appendix.

\section{OBSERVATIONS AND DATA REDUCTION}\label{sec:DR}
\begin{table*}[t]
    \centering
    \caption{Observations of the Perseus Cluster\label{tab:obs}}
    \begin{tabular}{ccccc}
     \hline 
     \hline
     \\[-0.75em]
      &  & RA & Dec & Cleaned Exposure \\
         ObsID & Date & (deg) & (deg) &  (ks) \\\\[-0.75em]
         \hline
         \\[-0.5em]
         60061361002 & 2015/11/03 & 49.930 & 41.496 & 14.6
         \\
        90202046002 & 2017/02/01 & 49.946 & 41.497 & 27.5%
         \\
         90202046004 & 2017/02/04 & 49.921 & 41.488 & 28.2
         \\
      \hline
    \end{tabular}
\end{table*}

\begin{figure}
    \includegraphics[width=1.0\columnwidth]{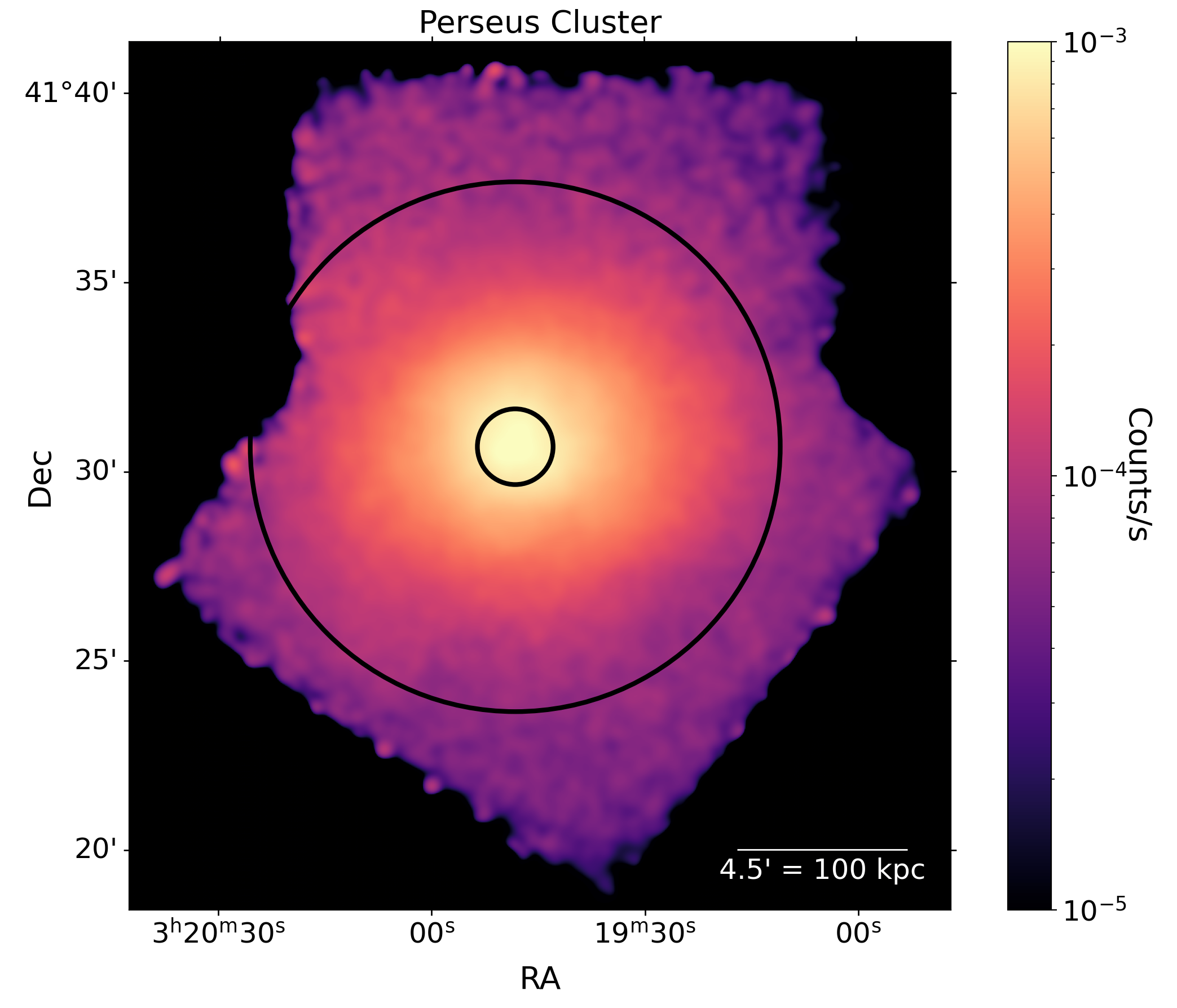}
    \caption{The combined background-subtracted, exposure-corrected, and AGN-subtracted image of Perseus in the 4--25~keV band, smoothed with a Gaussian ($\sigma = 4$~pixels).
    The black circles are the spectral extraction regions.
    \\}
    \label{fig:glob}
\end{figure}

\begin{figure*}
    \includegraphics[width=1.0\textwidth]{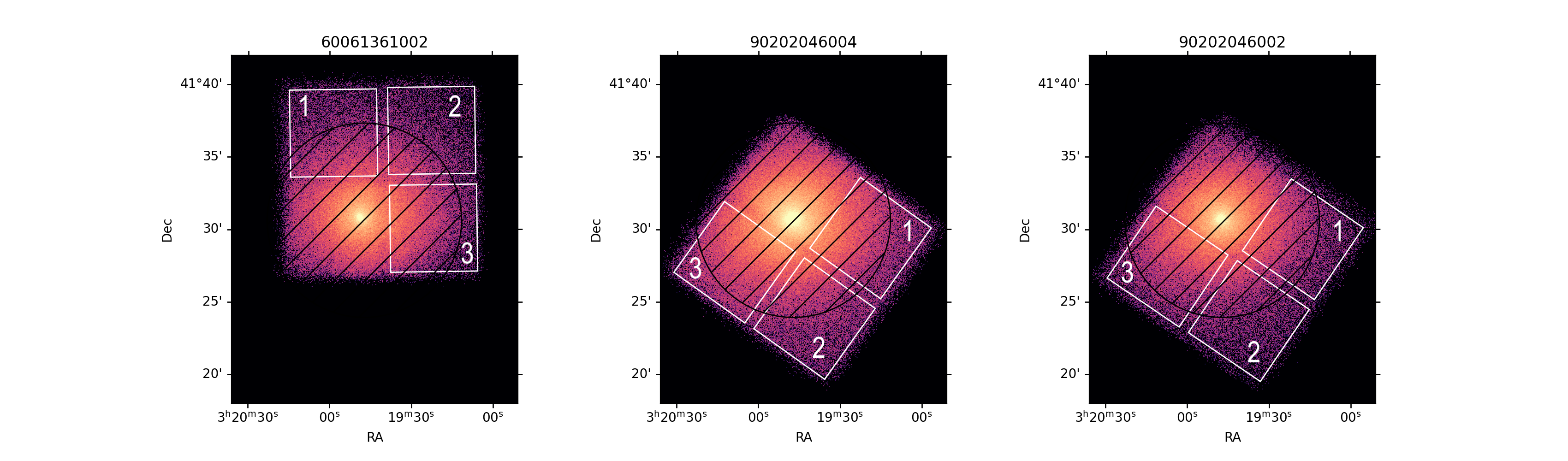}
    \caption{The combined FPMA and FPMB images from each of the three observations of the Perseus cluster. Shown in white are the extraction regions used to fit the background across the FOV with labels denoting the corresponding detector chip, and the black hashed circle is the exclusion region.}
    \label{fig:b_reg}
\end{figure*}

\begin{figure}
    \includegraphics[width=1.0\columnwidth]{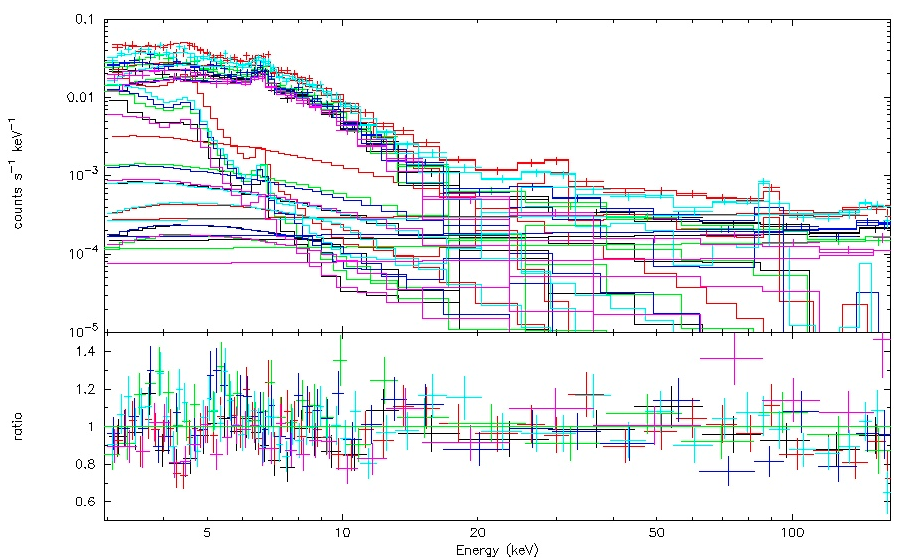}
    \caption{The background data and best-fit model for observation 90202046004. The spectra and models corresponding with FPMA detectors 1,2, and 3 (see Figure \ref{fig:b_reg}) are black, red, and green, respectively, and the three detectors on FPMB are blue, cyan, and magenta, respectively.}
    \label{fig:b_fit}
\end{figure}

In this work, we use data from three archival \textit{NuSTAR} observations of the Perseus cluster. Two of these observations (ObsIDs 90202046002 and 90202046004) are centered on 3C 84, the radio counterpart to the bright central galaxy of the cluster (NGC 1275). The third observation (ObsID 60061361002) is offset to the north by 2\farcm101. 
Together, these three observations yield 70.4 ks of data covering the central $r\lesssim 6\arcmin$ of the cluster. They are summarized in Table \ref{tab:obs} and their combined data are shown in Figure \ref{fig:glob}.

These data were processed using \textit{NuSTAR's} standard pipelines (HEASoft v6.30.1 and NUSTARDAS 1.9.7). 
From archival-provided cleaned event files, we manually screened data for periods of excess counts caused by high solar activity or high particle background, as can be caused by passages through the South Atlantic Anomaly.
Light curves in 1.6--20~keV (solar) and 50--160~keV (particle) bands with 100~s bins were constructed and periods with excess count rates were manually identified and removed from the Good Time Interval (GTI) files.
This process was repeated within the solar and particle bands for light-curves within the central $r\lesssim 6\arcmin$ of the cluster.
\textit{NuSTAR}'s stable background allows this simple procedure to quickly identify periods of high rates, which appear as clear outliers relative to typical variations during an observation.
This method typically removes less exposure time than conservative filtering with the {\tt SAAMODE=STRICT} and {\tt TENTACLE=yes} keywords set, which removes GTIs based on \textit{NuSTAR}'s geographic location and only catches periods of high particle background.
Additionally, this manual process improves the quality of GTI filtering; high background periods occasionally fall outside of the defined geographic regions excluded by the conservative filtering method, and in the opposite case, the background may be at nominal levels even while the telescope passes through those geographic regions.
The new GTIs were used to reprocess the data and create the cleaned event files with {\tt nupipeline}, which have been used to create all further data products.
The resulting exposure times are shown in Table \ref{tab:obs}.

Using \texttt{nuproducts}, we extracted higher level products from the cleaned event files.
Vignetted exposure maps were generated at 10 keV using \texttt{nuexpomap} in order to produce exposure-corrected 4--25~keV band images (e.g., Fig.~\ref{fig:glob}). 

To minimize contamination from the central AGN and isolate the central cool core from the hotter gas at larger radii, higher level data products--including Auxillary Response files (ARFs), spectra, and light curves--were extracted from a circular region ($r < 1\arcmin$) and a concentric annulus spanning the cluster gas ($1\arcmin < r < 7\arcmin$; see Figure \ref{fig:glob}). 
The spectra from both focal plane modules (FPMA and FPMB) of all three observations were jointly analyzed for this study.

\subsection{Background treatment}
\label{sec:bgd}

\textit{NuSTAR's} background spectrum can be separated into four distinct components that are each constrained within a reconstructed uncertainty. These components are described in detail by \citet{DanBullet}. Here, we briefly describe the components and list their 1$\sigma$ uncertainties.

The \textbf{internal background} ($\sigma_{\rm sys} = 3\%$) originates from two sources: fluorescence (22--32 keV) and activation (all energies, most prominent above 30 keV) lines within the telescope, and a flat component that represents instrumental noise and unresolved lines. This component dominates at energies $E > 20$~keV. 

The \textbf{aperture stray light} ($\sigma_{\rm sys} = 8\%$) component is a byproduct of \textit{NuSTAR's} open mast that separates the optics from the focal plane. 
Light from the cosmic X-ray background (CXB) leaks through the open mast and creates a gradient across the FOV. This source of background becomes obvious when a bright source 1--3\arcdeg\ from This component's uncertainty was constructed by simulating realistic CXB point sources across the detector plane, and \cite{DanBullet} finds that the CXB normalization values agree across the FOV within $8 \%$.

The normalization of this model should be within $\sim$10\% of the expected CXB level. 
Outside this range, the model is unphysical and is being confused with other background, foreground, or source components. 
When the background for observations 60061361002 and 90202046004 were being fit, the free aperture CXB normalization dropped to $\sim$60\% of the expected flux, so we instead fixed it to the nominal value.
Because the cluster emission is brighter than the aperture CXB flux, this choice has little impact on ICM measurements.

The \textbf{focused cosmic X-ray background} ($\sigma_{\rm sys} \sim 40\%$) is the cumulative light of unresolved X-ray sources--primarily AGN--focused within the FOV. 
This component is relatively uniform across the FOV, although the number of sources in typical regions drives the large systematic uncertainty.
The variance for this component is derived empirically from the Extended Chandra Deep Field South (ECDFS), and then adjusted conservatively for the smaller solid angle and higher flux limit sampled by galaxy cluster observations.

Lastly, scattered light from the Sun (and potentially from other sources) forms the so-called \textbf{solar} ($\sigma_{\rm sys}=10\%$) background component.
\cite{DanBullet} obtains the uncertainty by fitting the background of the ECDFS survey, and they find that this component is accurately reconstructed with a standard deviation of $10 \%$ after statistical fluctuations are accounted for. This same exercise is used to find the uncertainty of the instrumental background discussed above.

In order to fit these background components to our data, spectra were extracted from the regions shown in Figure~\ref{fig:b_reg}. 
Even though we used exclusion regions to block most of the cluster gas, the extracted spectra are still contaminated by significant thermal emission from the ICM. 
In order to account for this, we include an \texttt{apec} component \citep{Smith01_APEC} in fits to the spectra.
As a demonstration, the fitted background for observation 90202046004 is shown in Figure~\ref{fig:b_fit}. 

At energies $\lesssim 15$~keV, the cluster emission far dominates the background across the entire field of view, so the background model is not well constrained and residuals are left behind.
These residuals are likely due to imperfect modeling of background and cluster lines, which may suffer from small gain shifts (see Section~\ref{sec:spec_extraction:specmodeling}).
These residuals are not particularly concerning: Because the thermal emission far dominates in this energy band, a slightly imperfect background at $\lesssim 15$~keV will have negligible effects on model fitting.

Near $\sim$20~keV, the background becomes the dominant source of emission and it can be constrained more accurately with the background model.
Just before this background-dominated regime---near $\sim$15~keV---there is a slight excess that cannot be reasonably fit by any components in the current background model. 
This excess will be explored in greater detail in Section \ref{sec:analysis}.

Figure~\ref{fig:glob} shows the background-subtracted, exposure-corrected image combining all three observations from both focal plane modules. 
The AGN---which acts as a point source at the center of the cluster---has been subtracted using PSF models from \textsc{nucrossarf}, which will be discussed in more detail in Section~\ref{sec:analysis}.
Aside from the AGN, no point sources were apparent in soft (3--20 keV) or hard-band (20--50 keV) images of the cluster.
There was also no evidence of stray light leaking into the FOV from bright sources 1--3\arcdeg~off-axis, and the \textit{NuSTAR} stray-light evaluation tool reported no issues.
    
\section{Analysis}\label{sec:analysis}
\subsection{Spectral analysis and results}
\label{sec:spec_extraction}

\begin{figure*}[t]
\begin{center}
    \includegraphics[width=1.0\textwidth]{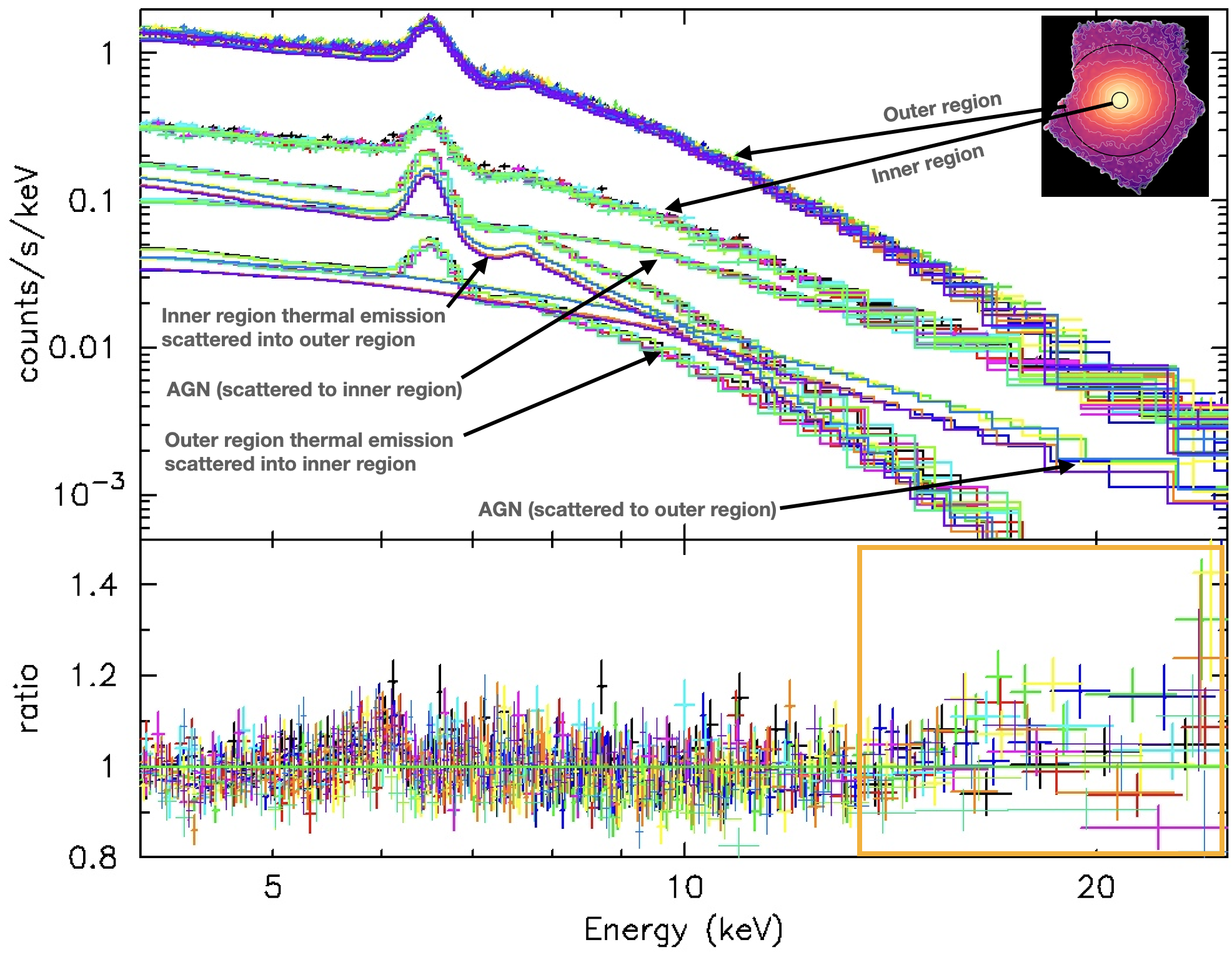}
    \caption{Single temperature fits to spectra from the two regions shown (inset), including an power law AGN model at the center of the inner region. \textbf{Top:} The data (crosses, with a different color for each individual spectrum) plotted with the different components of the model (solid lines, color-matched to the corresponding data) with components labeled. The higher-flux data belongs to the global outer annulus, and the lower flux data belong to the inner circular region (Fig.~\ref{fig:glob}). \textbf{Bottom:} The ratio of the data compared to the total model. The yellow box indicates the appearance of the hard excess.}
    \label{fig:1apec}
\end{center}
\end{figure*}

\textsc{nucrossarf}\footnote{\url{https://github.com/danielrwik/nucrossarf}} is a software suite that generates auxiliary response files (ARFs) to model cross-talk between different sources of emission in all regions of a \textit{NuSTAR} observation. 
This accounts for \textit{NuSTAR}'s large PSF wings, which cause contamination between sources (see Appendix~\ref{CR systematics} for more details). \citet{Tumer23} explains this set of IDL routines in detail, and we followed their procedures to extract spectra of the AGN emission in both an inner circular region and a global annulus encompassing the cluster gas (see Figure \ref{fig:glob}).
The AGN is modeled as a point source, while diffuse emission from these two regions (a circular region with $r < 1$\arcmin\ and a concentric annulus with $1\arcmin < r < 7\arcmin$) is modeled as extended sources following the spatial distribution of emission from a \textit{NuSTAR} image. 
The data were grouped using \textsc{grppha} to a minimum of three counts per bin.
Using \textsc{PyXspec} v2.1.1, we fit the spectra from 4--25~keV using Cash statistics \citep[C-stat;][]{cash79}. 
We did not include the 3--4 keV band---to avoid noise from the solar background---or data above 25 keV where the source falls below the systematic limit of the modeled background.
An example fit is shown in Figure \ref{fig:1apec}, where the data are additionally binned for display purposes only ($S/N = 10$, $N_{\rm bins} \leq 50$). 
The contribution from the AGN is modeled with a simple power law spectrum, and the thermal emission from both extraction regions is described by single-temperature \textsc{apec} models.
Some residuals appear near the iron line complex at 6~keV, and above $\sim15~$keV (yellow box), a subtle excess is apparent. 
As discussed in Section~\ref{subsec:models}, we find that this excess is statistically significant.

\subsection{Spectral modeling}
\label{sec:spec_extraction:specmodeling}

This analysis follows a similar process to past searches for IC \citep[e.g.,][]{DanBullet,DanComa,RB21,RB23,Tumer23}. 
To search for non-thermal contribution to the spectra, we first characterize the dominant thermal emission by fitting the data with \textsc{XSPEC}'s \textsc{apec} model, which models single-temperature (1T) emission.
Then, to asses the possible contribution from a non-thermal component describing IC emission, we introduce a \textsc{powerlaw} component (1T+IC). 
To ensure the 1T+IC model isn't just representing multi-temperature emission, we also consider
a two-temperature model (2T), which can adequately describe multi-temperature ICM emission and is well modeled by two \textsc{apec} components. 
We adopt these three models (1T, 1T+IC, 2T) for the following analysis. 

All of these models are convolved with by Galactic absorption (\textsc{wabs}), a gain response (\textsc{gain}), and normalization constants (\textsc{constant}). 
The Galactic absorption was modeled by a \textsc{wabs}\footnote{The \textsc{wabs} model is obsolete and should be replaced with the \textsc{tbabs} in future work. However, the effect of absorption on $E>3$~keV \textit{NuSTAR} ICM spectra is negligible and does not affect the results presented here.} component that was fixed to $n_H = 1.4 \times 10^{21}~\text{cm}^{-2}$, matching \cite{Rani18} (calculated using the nH FTOOL from HEASoft).
Given that we only include events with energies $E>4$~keV, the exact amount of the absorption and how it is modeled is negligible and could have been completely ignored.
The gain response accounts for the slight offset between actual photon energy and channel energy, which is necessary to achieve a consistent redshift with past measurements; the redshift is left fixed to this value (see Section~\ref{sec:intro}).
We fixed the gain slope to $1.0$, while the offset was left as a free parameter as in \citet{RB23}. 
The normalization constant allows the model to accommodate for any flux or instrumental differences between the three observations and instruments (FPMA and FPMB). 
Unless explicitly stated otherwise, Galactic absorption, gain response, and normalization constants were included for the entirety of our analysis.

\subsection{Modeling the AGN}

In order to model the AGN contribution to the extracted spectra, we add a point source model at the AGN's location in \textsc{nucrossarf}. 
Assuming that the AGN emission is point-like, \textsc{nucrossarf} generates ARFs that model the expected counts from the AGN in our two regions, accounting for the energy-dependent scatter of the PSF.
The ARF for the smaller central region is essentially identical to what would be produced by {\tt nuproducts}, and the ''cross-ARF'' for the larger annulus accounts for the emission scattered into that region by the PSF wings.
These ARFs are used to convolve a \textsc{powerlaw} spectrum to model the AGN counts in each region.
Given perfectly-realized response functions and a non-variable AGN flux over time, the power law index and normalization would then be applied across both telescopes and the three observations; to allow for time variability and imperfect energy-independent calibration, we allow the cross-calibration constant to float between telescopes and observations (but not between the regions of a single dataset).
Linking the photon index across datasets, we find a best-fit value of $\Gamma = 1.82^{+0.13}_{-0.01}$, which is consistent with \citet{Rani18}, who finds $\Gamma = 1.85^{+0.12}_{-0.14}$ (observation 90202046002) and $\Gamma = 1.90^{+0.13}_{-0.11}$ (observation 90202046004). 
However, the C-statistic space was not entirely well-behaved, especially on the lower end of the parameter range.
In an effort to minimize degeneracy with the thermal cluster component, we tried changing the region size from $r<1\arcmin$ to $r<30\arcsec$, but this did not improve the behavior, demonstrating the difficulty of simultaneously fitting the AGN and cluster emission without other constraints or assumptions.
We only aim to accurately model the AGN emission so that it does not bias ICM measurements, so in light of this issue, we fixed the powerlaw index to $\Gamma = 1.82$ for the remainder of the analysis.
This value is consistent with \citet{Rani18}, whose focus was on the AGN, thus supporting our choice.
Within the outer region, we find that scattered light from the AGN constitutes $\sim 26 \%$ of the 20-25~keV emission and $\sim 4 \%$ of the 4-25~keV emission in our ICM spectra.


\begin{figure*}[t]
\begin{center}
    \includegraphics[width=0.9\textwidth]{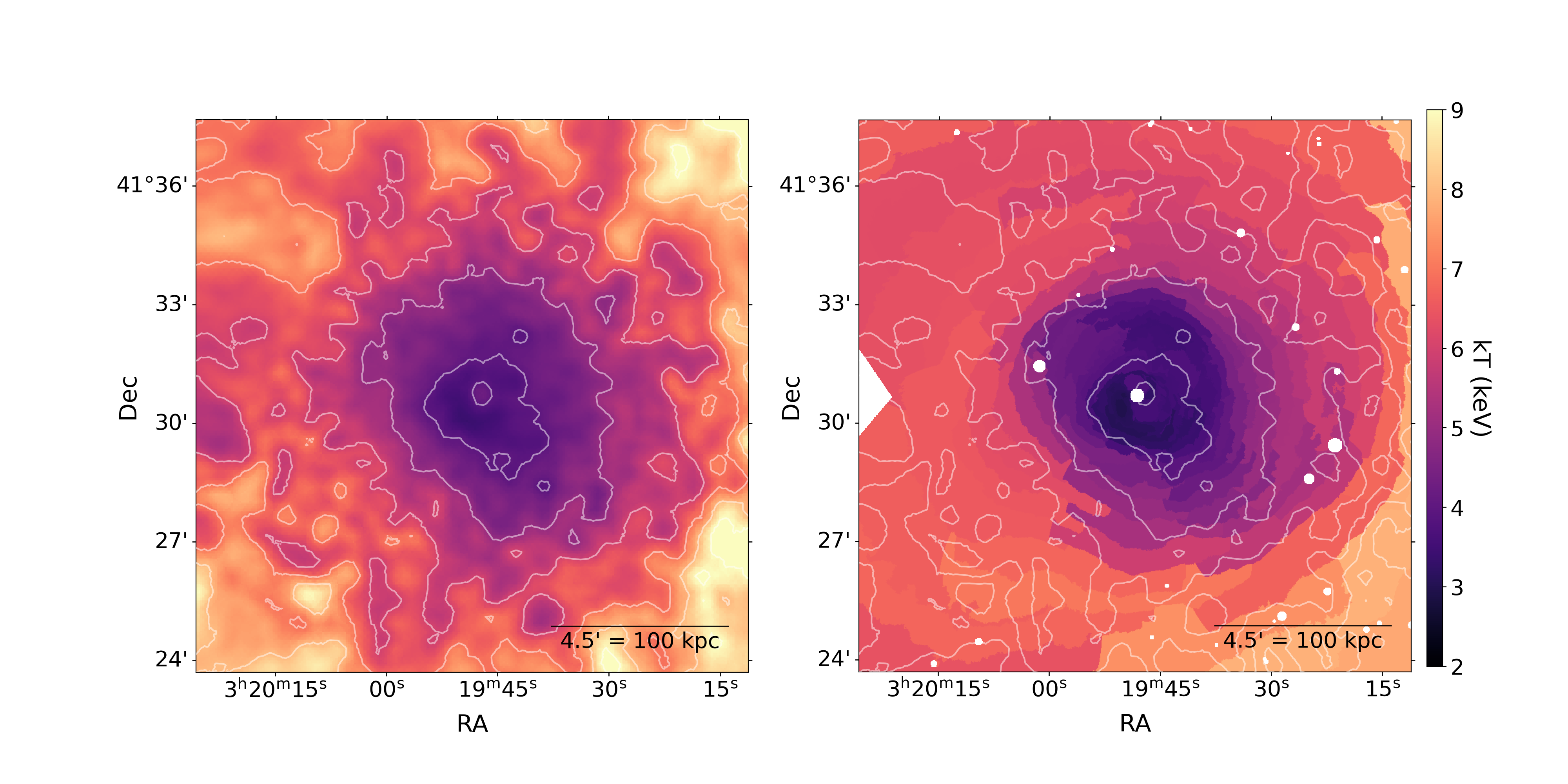}
    \caption{Comparison between the temperature map produced using \textit{NuSTAR} data (\textit{left}) and Chandra data \citep[\textit{right,} from][]{Pers_Chand_Sanders07}. \textit{NuSTAR} temperature contours are overlaid in both plots for easy comparison.}
    \label{fig:tmap}
\end{center}
\end{figure*}

\begin{figure*}
\begin{center}
    \includegraphics[width=0.8\textwidth]{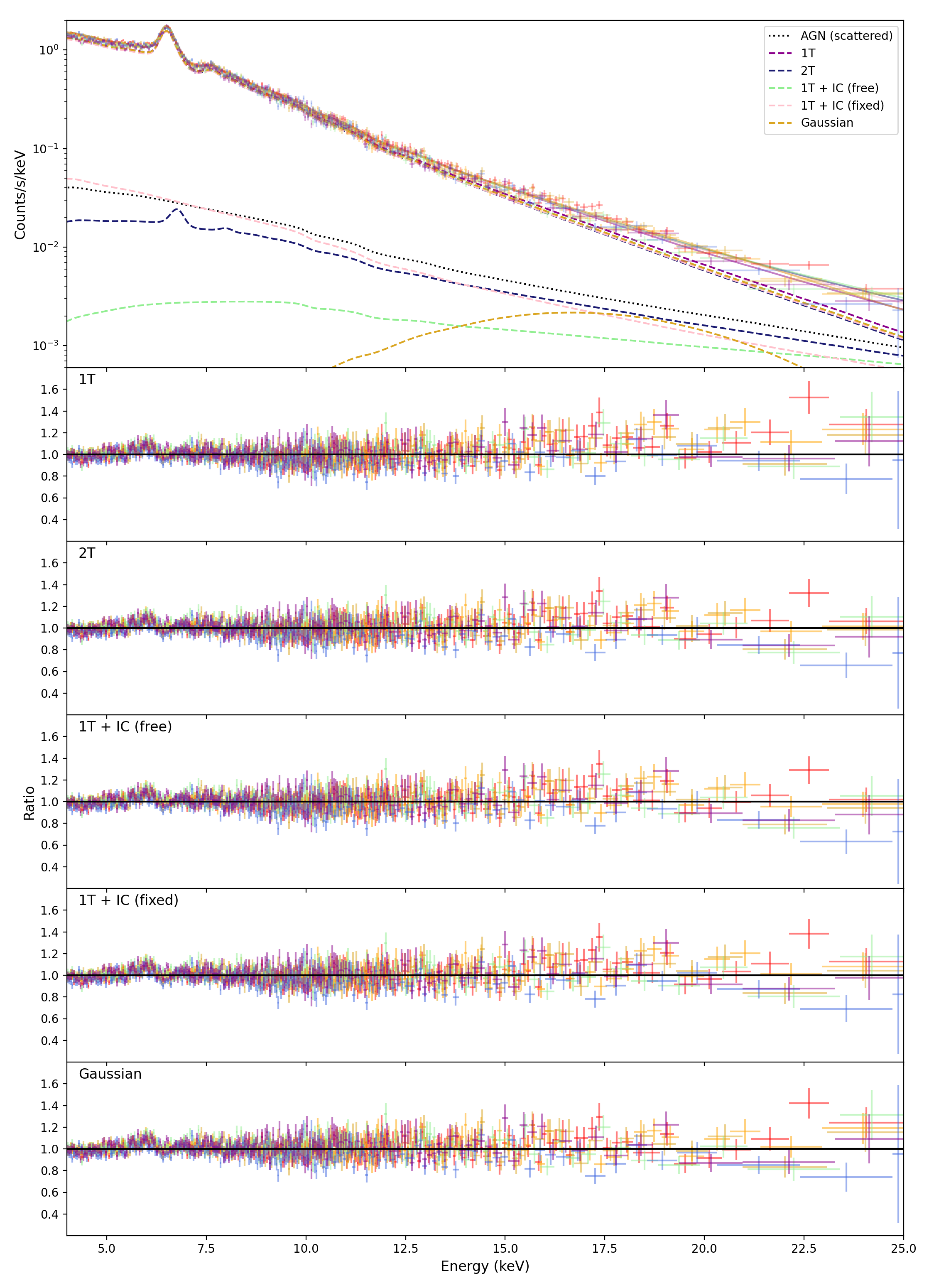}
    \caption{\textbf{Top panel:} Comparison of the components of four different models that could describe the hard excess in the Perseus cluster (solid lines are the total model, and the dashed lines of the same color are the individual components of the model): 1T, 2T, 1T + IC (free photon index), 1T + IC (fixed photon index), and a Gaussian. The total model is plotted by solid lines while components are dashed. The dotted black line gives the AGN's contribution to the model. \textbf{Bottom five panels:} the ratio of the models to the data, demonstrating the goodness of fit across the energy range. The C-stats and best-fit parameters are shown in Table \ref{tab:pars}}
    \label{fig:model_comp}
\end{center}
\end{figure*}

\subsection{Modeling the ICM}\label{subsec:models}

Our goal is to describe the nature of the emission in the outer ($1\arcmin < r < 7\arcmin$) region that dominates the data and is less contaminated by the central AGN.
For the central region, we model the gas with an independent single temperature thermal component and, when applicable, non-thermal components with independent normalizations but with indices tied to that in the outer region.
This procedure allows for extensive flexibility in modeling the inner region spectrum.
While it reduces the constraining power of the inner region models, this procedure also removes as much as possible constraints on model components that describe the outer region spectrum.
This conservative approach allows the data to constrain our models as realistically and accurately as possible.
In the following, we focus on the model parameters found for the outer region, treating the inner region values more like nuisance parameters.

\subsubsection{Single-Temperature Model}

Up to $\sim 15~$keV, the global spectrum of Perseus's ICM is adequately described by a single temperature (1T) model, with C-stat/Degrees of Freedom (hereafter C-stat/$\nu$) = 6537.35/5983.
However, this model assumes that the ICM has a uniform temperature, which is only a first approach to the system.
Following \cite{Potter23}, we obtain \textit{NuSTAR} and \textit{Chandra} temperature maps (Figure \ref{fig:tmap}) by fitting single-\textsc{apec} models to coarse spectra that are created by extracting background-subtracted, exposure-corrected counts from three narrow energy bands ($3-5$, $6-10$, and $10-20$~keV) in regions ($r\geq~5$ pixels) spaced every fifth pixel across the FOV.
The center of each region is assigned its corresponding best-fit temperature, and the resulting grid of temperatures is interpolated to produce an image. 
The maps in Figure \ref{fig:tmap} hint to a considerable amount of temperature variation, and a 1T model, only able to estimate an average temperature in the region, may be too simplistic.

Figure \ref{fig:1apec} shows the 1T fit to the data, and residuals appear around the iron line. 
We have seen such features in other \textit{NuSTAR} observations of relaxed clusters;
it may be that line abundances at this location in the spectrum are not well captured by our assumptions---single temperature, solar abundance for all elements---or it may indicate some very small calibration issue with the response around the Fe complex.
The residuals cannot be attributed to problems with redshift or abundance, and given how weak and localized they are in the spectrum, we do not expect them to affect our broadband analysis.

Above $15$ keV, a slight ($\sim10\%$) excess becomes apparent (see Figure \ref{fig:1apec} and \ref{fig:model_comp}). We explore this hard excess in more depth by testing several different models.

\subsubsection{Two-Temperature Model}

As shown in Figure \ref{fig:tmap}, the Perseus Cluster has a multi-temperature structure that should not be fully consistent with a 1T model.
As explained in \citet{RB21}, the thermal continua described by \textsc{APEC} models have a lack of strong features at high energies that ultimately allows multi-temperature structures to reduce to a two-temperature (2T) scenario.

While a 2T model describes the data with a lower fit statistic than the 1T model (C-stat/$\nu$ = 6482.52/5981), the best fit parameters for the second \textsc{APEC} component are unconstrained, and its temperature reaches the default upper limit of 64~keV. 
At these high temperatures, the turn-over of the bremsstrahlung is not visible within the analyzed energy band, so it resembles a powerlaw describing a non-thermal emission mechanism.
Additionally, 64~keV is not a physically realistic temperature for the ICM, especially in a relaxed, cool-core cluster like Perseus.
Hence, the hard excess does not describe a physical thermal component.

When a physically-motivated upper limit of $kT \leq 9$ is applied to match the highest temperature in the global region (seen in the \textit{NuSTAR} temperature map in Figure \ref{fig:tmap}), the C-stat/$\nu$ becomes 6497.29/5981.

\subsubsection{Non-Thermal Model}
\label{sec:Non-Thermal}

In order to assess the contribution of possible IC emission, we included a powerlaw component to account for possible non-thermal processes (1T + IC). 
The powerlaw (C-stat/$\nu$ = 6475.78/5981)  was strongly preferred over a 1T (C-stat/$\nu$=6537.35/5983) or 2T (C-stat/$\nu$ = 6482.52/5981) model.
However, its best-fit results are inconsistent with the expectations from a diffuse IC component.
When left free to vary, the best-fit photon index ($\Gamma$) dropped to 0.254.
This indicates that the excess emission -- if described by a powerlaw -- is extremely flat. 
\citet{GM21} found that the radio index of the Perseus Cluster to be $\alpha = 1.2$, which gives an expected IC photon index $\Gamma_\text{IC} = \alpha + 1 = 2.2$.
When the photon index is fixed to 2.2 (1T + IC), the resulting statistics (C-stat/$\nu = 6498.18/5981$) is improved with respect to a 1T model, but it is worse than either a 2T model or a 1T + IC model with a free $\Gamma$ parameter.

When $\Gamma$ is fixed to the expected value of 2.2, we obtain an upper limit on the $4-25$~keV IC flux of $1.5 \times 10^{-11}$~erg~s$^{-1}$~cm$^{-2}$.
As discussed in section \ref{subsec:B_calc}, this constrains the magnetic field to be  $B \geq 0.35~\mu G$.

\subsubsection{Testing other possibilities}

None of the standard models discussed thus far (1T, 2T, and 1T
+IC) can both adequately describe the data and be physically motivated. To further investigate the origin of the hard excess, we tested three additional models.

\noindent
\textbf{1T * Comptonization:} 
Some plasmas may be able to Comptonize their own thermal bremsstrahlung emission, which could produce a hard excess. 
This phenomenon is modeled by convolving an \textsc{apec} with \textsc{XSPEC}'s \textsc{simpl} model \citep[][]{Steiner09_simpl}.
While we don't expect to see this phenomenon to occur in the diffuse ICM, a Comptonization model can mimic a non-Maxwellian scenario. 

In models of collisionally ionized plasma, the base simplifying assumption is that collisions result in a Maxwellian distribution of electrons. 
However, simulations by \citet{NonMax_GC_Mergers_Akahori10} show that cluster mergers can disturb the thermal equilibrium of the plasma.
Relaxed clusters such as Perseus are not known to exhibit non-Maxwellian behavior, but we tested whether a similar model would be able to explain our excess. 

We applied this Comptonization model to both the inner and outer regions of diffuse emission. We fixed the \texttt{UpScOnly} parameter to 0, which specifies that photons are only up-scattered. This choice reflects the fact that the relativistic electrons responsible for the Comptonization in this scenario are at much higher energies than the thermal photons they are scattering, and so we approximate that no photons are down-scattered. 
We linked the other parameters (the fraction of scattered photons and $\Gamma$, which is the photon index of the scattered emission) across both regions. 
With C-stat/$\nu$~=~6477.73/5982, this self-Comptonization model slightly improved the statistics with respect to the 2T model, but it was disfavored over 1T+IC with a free $\Gamma$.

\noindent
\textbf{Non-Maxwellian:}
Because the Comptonization model---which mimics a non-Maxwellian scenario--- fit the data fairly well, we implemented a more physically accurate non-Maxwellian model in search of a more realistic description of the data.
\href{https://github.com/AtomDB/kappa}{PyKappa}\footnote{Find on GitHub: https://github.com/AtomDB/kappa}, which can be imported into \textsc{PyXSPEC} as a custom model, implements the findings of \cite{NonMax_Fe_Hahn15} to model a non-Maxwellian equivalent of the \textsc{APEC}. 

Fitting PyKappa to data is computationally expensive, so we fit a single \textsc{apec} to the central thermal region and a \textsc{kappa} model to the global annulus excluding the inner circle (Figure \ref{fig:glob}) in order to speed up fitting. 

\movetabledown=65mm
\begin{rotatetable*}
\begin{deluxetable*}{l|c|cccc|cccccc|c}
    \tablecaption{Best-Fit Model Parameters\label{tab:pars}}
    \startdata
    \small
         & \textbf{AGN}\tablenotemark{g} & 
         \multicolumn{4}{c|}{\textbf{ICM$_\text{inner}$}} &
         \multicolumn{6}{c|}{\textbf{ICM$_\text{outer}$}} &
         \\
         & 
         & 
         kT & 
         & 
         &
         $n_\text{excess}$ &
         kT & 
         & 
         &
         kT/$\Gamma$/E & $Z$\tablenotemark{a}/F$_\text{sctr}$/$\sigma$/$\kappa$\tablenotemark{b} & 
         $n_\text{excess}$ &
         \\
          \textbf{Model}
          &
          $\kappa$\tablenotemark{c} &
          (keV) & 
          $Z$\tablenotemark{a} &
          $n$\tablenotemark{d} & 
          (-\tablenotemark{c}/-\tablenotemark{e}) & 
         (keV) & 
          $Z$\tablenotemark{a} & 
          $n$\tablenotemark{d} & 
          (keV/-/keV) & 
          (-/-/keV/-) & 
          (-\tablenotemark{d}/-\tablenotemark{c}/-\tablenotemark{a}) &
          \textbf{C/$\nu$}
         \\ \hline
         \renewcommand{\arraystretch}{1.2}
         \textbf{1T} & 
          9.92$^{+0.24}_{-0.39}$ &
          3.44$^{+0.12}_{-0.09}$ & 0.392$^{+0.031}_{-0.031}$ & 110$^{+2}_{-2}$ & - & 
          5.45$^{+0.04}_{-0.03}$ & 0.345$^{+0.010}_{-0.008}$ & 439$^{+6}_{-2}$ & - & - & - &
          \textbf{6537.35/}
          \\
          &
          & & & &
          & & & & & & &
          \textbf{5983}
         \\ \hline
        \textbf{2T} & 
         9.65$^{+0.43}_{-0.41}$ & 
         3.57$^{+0.19}_{-0.20}$ & 0.382$^{+0.038}_{-0.035}$ & 106$^{+7}_{-6}$ & - & 
         5.22$^{+0.10}_{-0.18}$ & 0.342$^{+0.013}_{-0.016}$ & 446$^{+7}_{-7}$ & 
         $\geq 7.0$ & *\tablenotemark{f}  & 3.62$^{+25.58}_{-1.45}$ &
          \textbf{6482.52/}
         \\
         &
          & & & &
          & & & & & & &
          \textbf{5981} 
          \\ \hline
        \textbf{1T + IC} & 
          8.45$^{+1.48}_{-4.13}$ & 
          3.76$^{+0.37}_{-0.32}$ &  0.355$^{+0.050}_{-0.055}$ & 108$^{+11}_{-7}$ & 0.014$^{+1.502}_{-0.009}$ & 
          5.30$^{+0.08}_{-0.11}$ & 0.343$^{+0.012}_{-0.011}$ & 445$^{+7.6}_{-7.5}$ & $\leq$~1.5 & - & 0.02$^{+0.98}_{-0.01}$ &
          \textbf{6475.78/}
          \\
          \textbf{(free $\Gamma$)}
          &
          & & & &
          & & & & & & &
          \textbf{5981} 
          \\ \hline
        \textbf{1T + IC} & 
          9.69$^{+0.20}_{-0.16}$ & 
          3.55$^{+0.04}_{-0.05}$ & 0.384$^{+0.020}_{-0.023}$ & 106$^{+2}_{-2}$ & 0.005$^{+0.600}_{-0.005}$ & 
          5.28$^{+0.25}_{-0.89}$ & 0.355$^{+0.007}_{-0.008}$ & 431$^{+3}_{-9}$ & 2.2 & - & 8.23$^{+0.86}_{-1.88}$ &
          \textbf{6498.18/}
          \\
          \textbf{(fixed $\Gamma$)}
          &
          & & & &
          & & & & & & &
          \textbf{5981} 
          \\ \hline
        \textbf{1T *} & 
          9.96$^{+93.44}_{-0.15}$& 
          2.87$^{+0.23}_{-0.03}$ & 0.614$^{+0.152}_{-0.037}$ &  117$^{+9}_{-2}$ & - & 
          4.90$^{+0.14}_{-0.02}$ & 0.448$^{+0.045}_{-0.007}$ & 440$^{+7}_{-2}$ & *\tablenotemark{f} & 0.26$^{+0.06}_{-0.01}$ & - &
          \textbf{6477.73/}
          \\
          \textbf{Compt.}
          &
          & & & &
          & & & & & & &
          \textbf{5982} 
          \\ \hline
         \textbf{1T +} & 
          9.15$^{+0.82}_{-1.80}$ & 
          3.66$^{+0.42}_{-0.27}$ &  0.372$^{+0.043}_{-0.047}$ &  106$^{+6}_{-6}$ & 0.040$^{+0.170}_{-0.040}$ & 
          5.29$^{+0.07}_{-0.07}$ & 0.343$^{+0.012}_{-0.011}$ & 448$^{+7}_{-7}$ & 17.7$^{+1.5}_{-1.4}$ & 3.60$^{+2.72}_{-1.27}$ & 0.14$^{+0.09}_{-0.06}$ &
          \textbf{6471.20/}
          \\
          \textbf{Gauss}
          &
          & & & &
          & & & & & & &
          \textbf{5980} 
          \\ \hline
         \textbf{PyKappa} & 
          9.70$^{+0.35}_{-0.24}$ & 
          3.54$^{+0.08}_{-0.07}$ &  0.389$^{+0.032}_{-0.032}$ &  105$^{+3}_{-2}$ & - & 
          5.56$^{+0.08}_{-0.02}$ &  0.270$^{+0.008}_{-0.006}$ & 606$^{+6}_{-7}$ & - & 23.5$^{+4.7}_{-5.1}$ & - &
          \textbf{6536.28/}
          \\
          &
          & & & &
          & & & & & & &
          \textbf{5983} 
          \\ \hline 
%
    \enddata
\tablenotetext{a}{Relative to solar abundance}
\tablenotetext{b}{Degree of non-Maxwellianness of the electrons}
\tablenotetext{c}{Normalization of the \textsc{powerlaw} model: $10^{-3}$~photons~keV$^{-1}$~cm$^{-2}$s$^{-1}$ at 1 keV)}
\tablenotetext{d}{Normalization of the \textsc{APEC} model: $\frac{10^{-14}}{4 \pi [D_A (1 + z)]^2}\int n_e N_H dV $~$10^{-3}$~cm$^{-5}$, where $dV$ is the volume element, $D_A$ is the angular diameter distance, $n_e$ is the electron density, and $n_H$ is the hydrogen density}
\tablenotetext{e}{Normalization of the \textsc{Gauss} model: total $10^{-3}$~photons~cm$^{-2}$s$^{-1}$ contained in the line}
\tablenotetext{f}{Unconstrained parameter.}
\tablenotetext{g}{$\Gamma$ fixed to 1.82 for all observations}
\end{deluxetable*}
\end{rotatetable*}

\begin{figure*}[t]
\begin{center}
\includegraphics[width=0.3\linewidth]{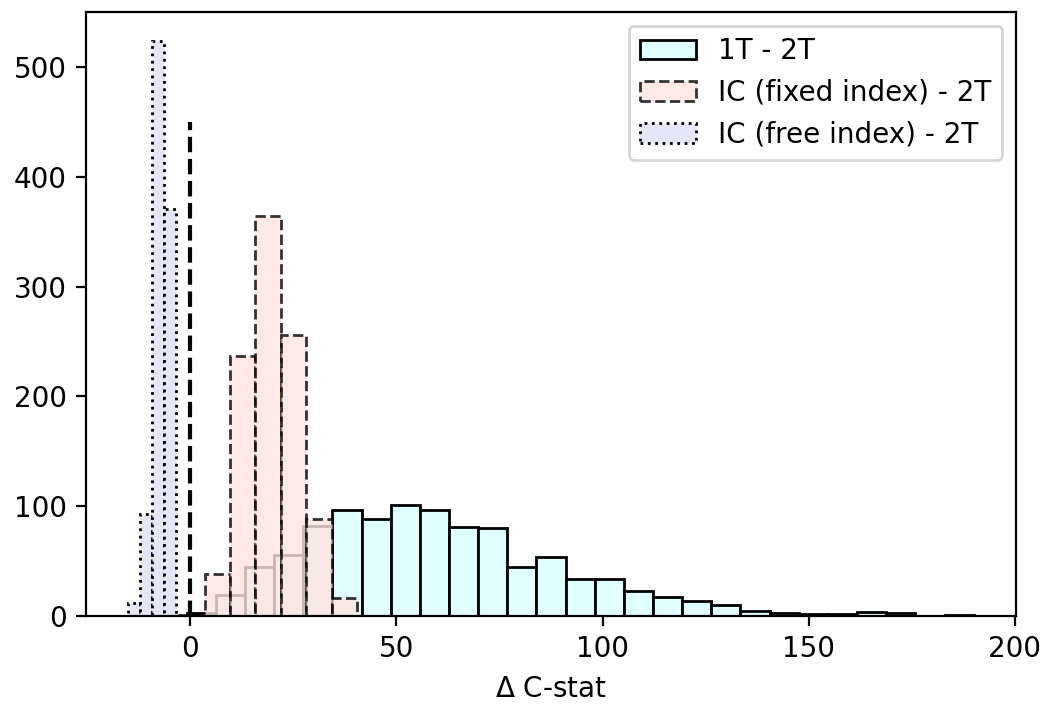}
\includegraphics[width=0.3\linewidth]{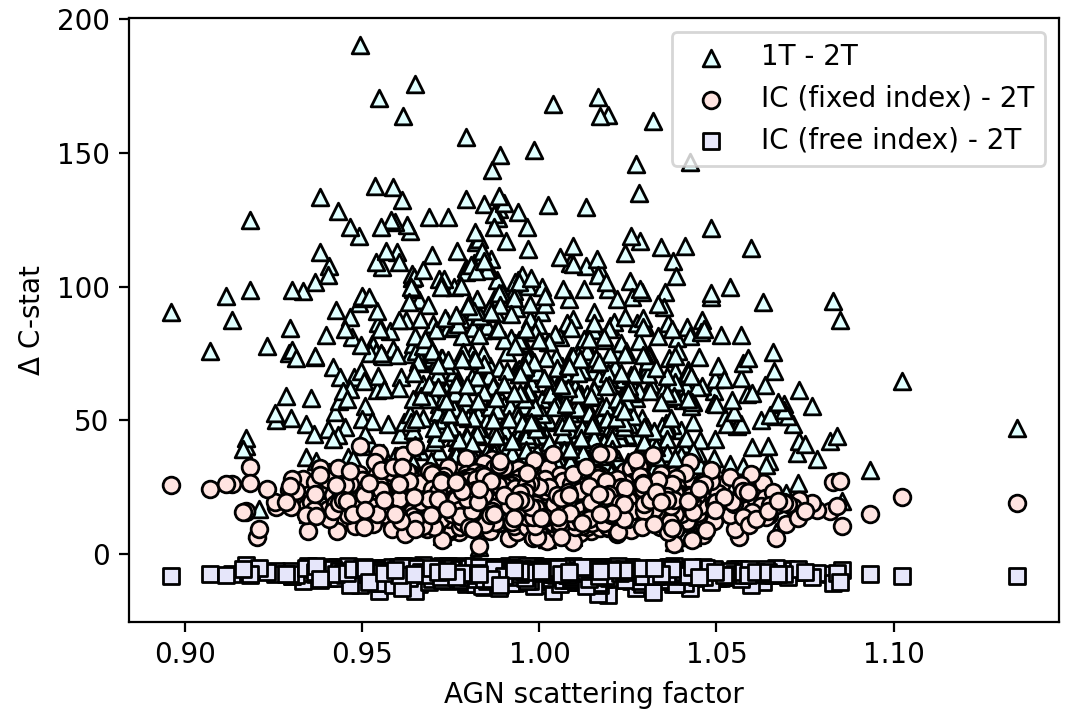}
\includegraphics[width=0.3\linewidth]{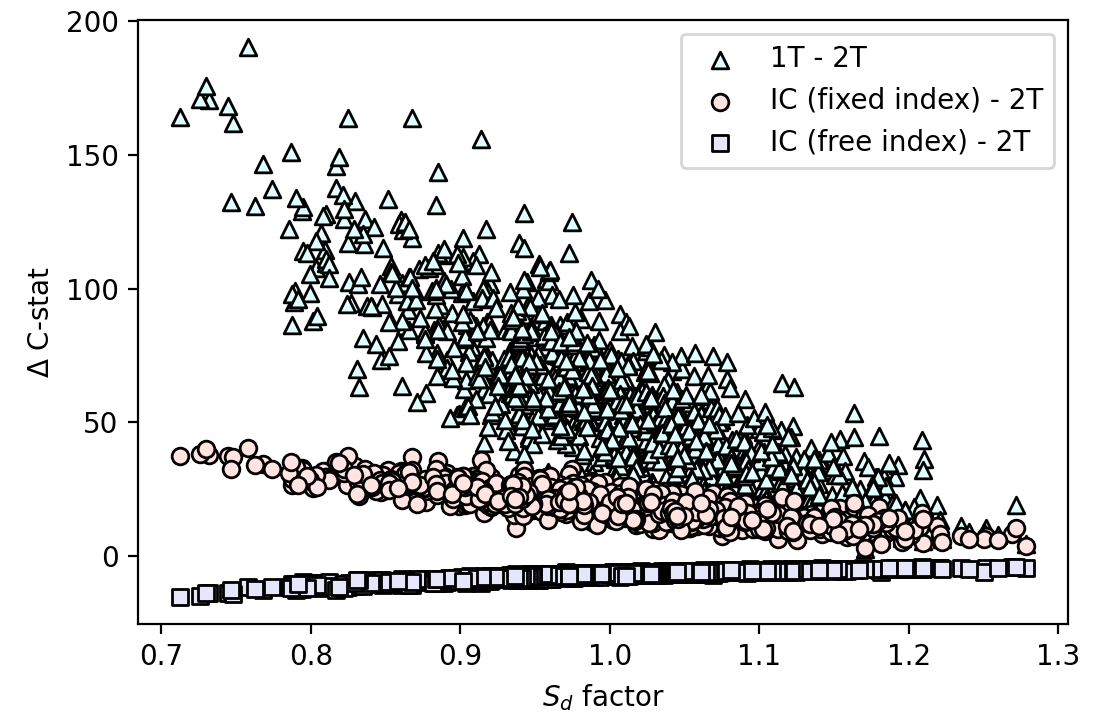}
\end{center}
    \caption{The difference in the C-statistic ($\Delta$~C-stat) between the 2T model and 3 other models: a 1T (cyan; triangles), an IC model with a fixed photon index (pink; circles), and an IC model with a free powerlaw index (purple; squares) for 1000 realizations of the background model and scattered AGN contribution.
    The 2T model is favored when $\Delta$~C-stat~$>0$. 
    \textbf{Left:} The histograms of $\Delta$~C-stat, where the dashed line marks $\Delta$~C-stat~$=0$.
    \textbf{Middle:} $\Delta$~C-stat as a function of the AGN scattering factor, which shifts the contribution of the AGN relative to the nominal value (the AGN contribution is strengthened for a factor $>1$). 
    \textbf{Right:} $\Delta$~C-stat as a function of the $S_d$ (aperture stray light) factor, which shifts the strength of the $S_d$ background component relative to its nominal value ($S_d$ is strengthened for a factor $>1$). See section \ref{sec:cr_comp} for discussion.}
    \label{fig:sys_2apec}
\end{figure*}

This model slightly improved the statistics with respect to the 1T model but was otherwise disfavored compared to other models, with a C-stat/$\nu$ $= 6536.28/5983$. The kappa parameter---which describes the degree of non-Maxwellianness \citep[see][]{Cui19_kappa}---had a value of 23.5$^{+4.7}_{-5.1}$, which implies only a slight deviation from Maxwellianness.

\noindent
\textbf{1T + Gaussian:} We find that the hard excess was best described by a wide Gaussian (\textsc{gauss}) added to the 1T model.
Applying the Gaussian to the model describing the AGN emission (\textsc{powerlaw} + \textsc{gauss}; C-stat/$\nu$~=~6527.02/5980) only slightly improves the overall fit compared to a 1T model. 
However, adding the Gaussian component to the model describing the ICM (\textsc{apec} + \textsc{gauss}, C-stat/$\nu$~=~6471.20/5980) made a significant improvement, and it was statistically favored over all of our other models. 
The Gaussian was initialized at 20~keV with a width of 1~keV to roughly match the observed properties of the hard excess.
After fitting, the centroid became 17.7 keV, and the width became 3.6~keV.

\subsection{Applying Crossarf and Background Systematics} 
\label{sec:cr_comp}

As described in section \ref{sec:spec_extraction}, we use \textsc{nucrossarf} to extract spectra and generate cross-ARFs for three sources: a point-source at the center of the cluster (representing the AGN), an extended source in a small ($r = 1$') circular region, and a second extended source in a concentric annulus (r$_\text{in}=1$' and r$_\text{out}=7$', see Figure \ref{fig:glob}).

In the Appendix, we find that \textsc{nucrossarf} is able to model PSF leakage from point sources with a $3.4 \%$ systematic uncertainty within similarly sized regions and for a similar bandpass.
This uncertainty governs how well we can model the AGN emission being scattered into the outer region, where it could be confused for diffuse non-thermal emission.
Because the modeled AGN light scattered into the outer region contributes $25\%$ of the total 20-25~keV ICM emission, it would need to be $39\%$ stronger than has been predicted by \texttt{nucrossarf} in order to explain the $10\%$ excess, which far exceeds the $3.4\%$ uncertainty.
However, to thoroughly account for this uncertainty in our models,
we followed a similar process that \citet{DanBullet} used to characterize the systematic error of the background components; we applied this process to the AGN cross-ARF as well as to the background spectra.

Following this method, we generated 1000 realizations of the AGN cross-ARF. We varied the normalization of the cross-ARFs with a scatter of 3.4$\%$.
Similarly, we also simultaneously generated 1000 realizations of the background model, assuming a Gaussian distribution (with $\sigma$ equal to the systematic error) about the nominal value for each component. 
For each iteration, we fit four of our models [1T, 2T, 1T+IC (free $\Gamma$), and 1T+IC (fixed $\Gamma$)] with the modified background and AGN cross-ARF, finding new values of C-stat for each model, that we compare relative to each other for each iteration.
The results are shown in Figure \ref{fig:sys_2apec}.

\begin{figure*}
    \begin{center}              
        \includegraphics[width=0.24\textwidth]{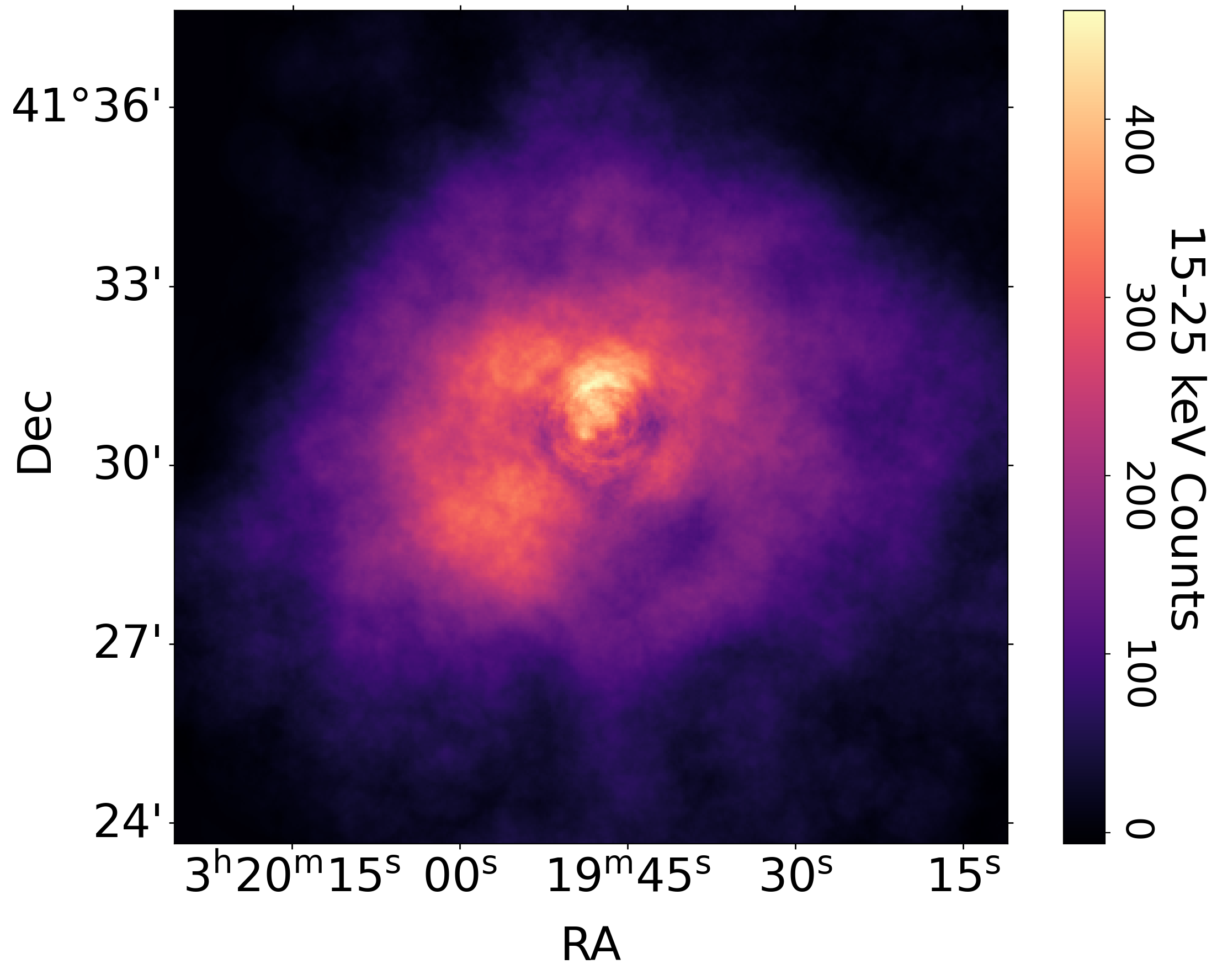}
        \includegraphics[width=0.24\textwidth]{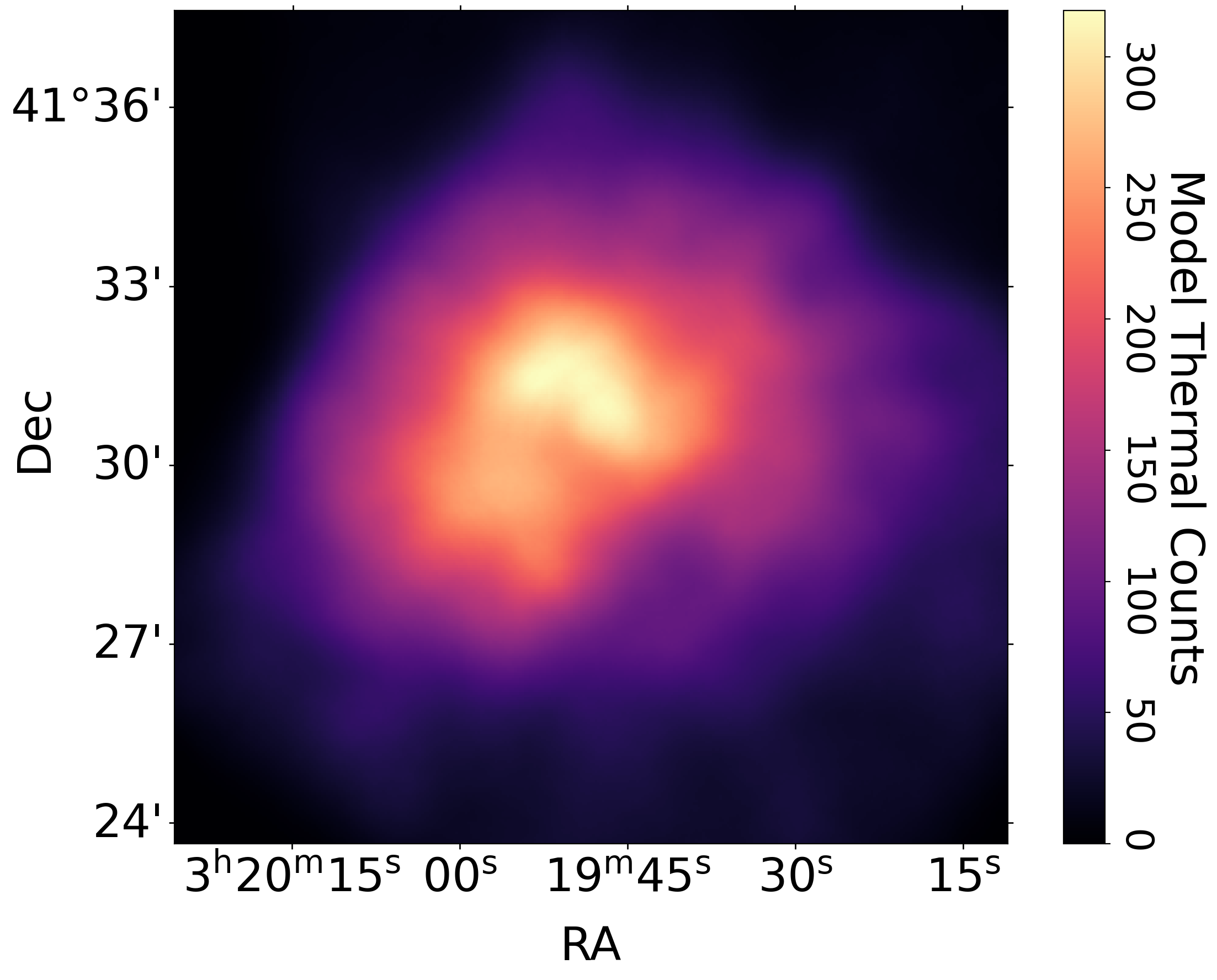}
        \includegraphics[width=0.24\textwidth]{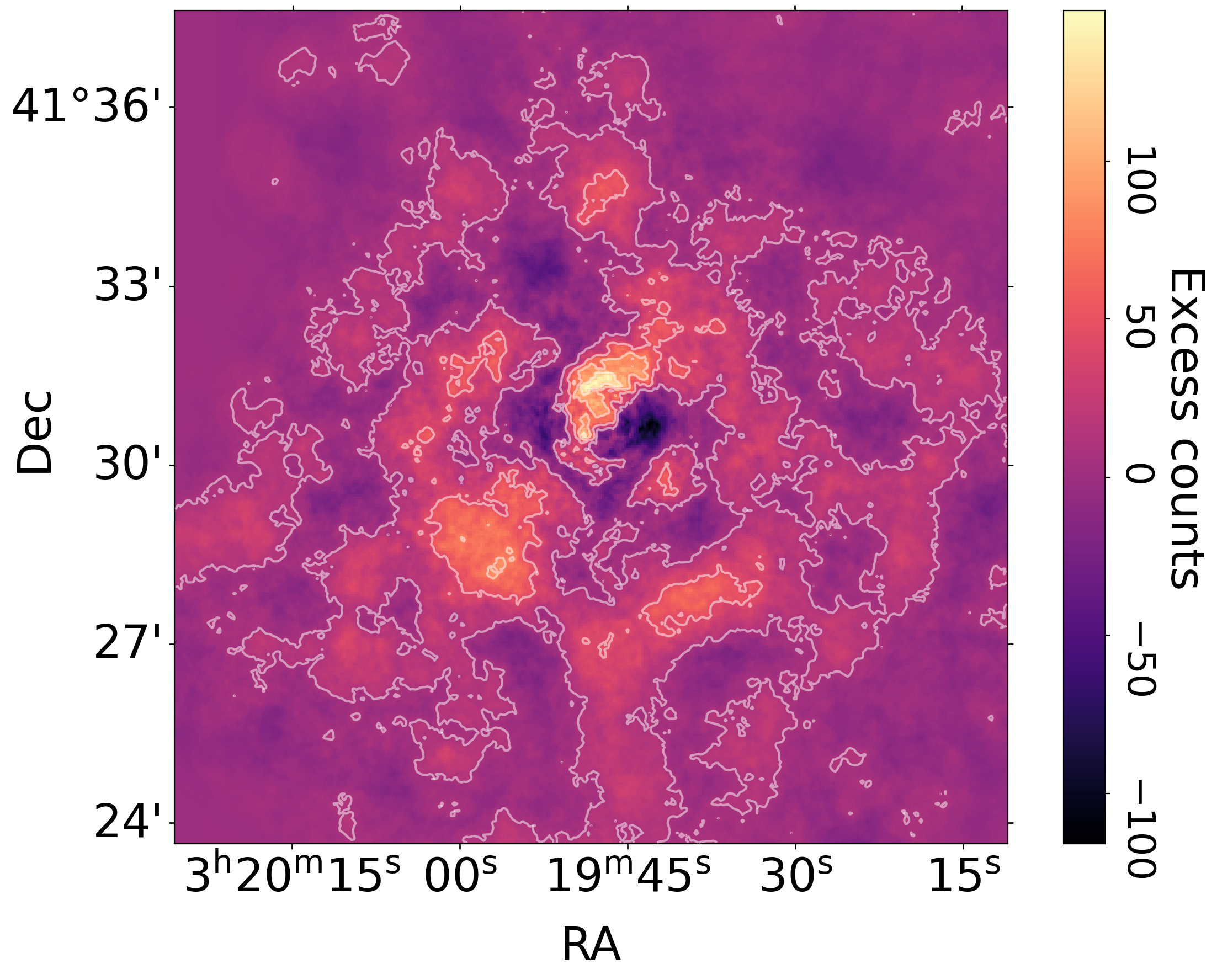}
        \includegraphics[width=0.24\textwidth]{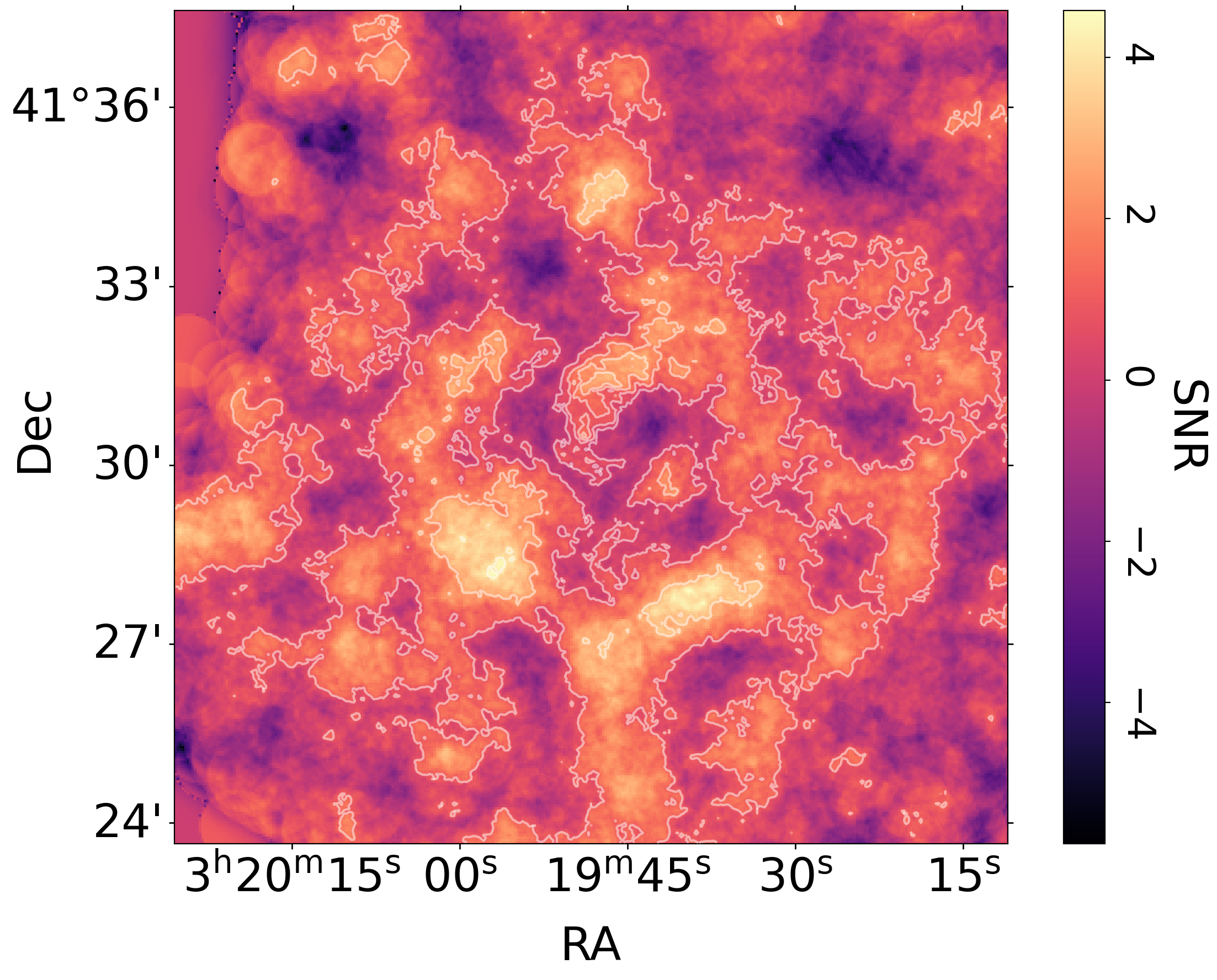}
    \end{center}
    \caption{\textbf{Left:} The AGN and background-subtracted counts from 15-25 keV for the three summed observations. \textbf{Middle-left:} The modeled thermal emission in the hard (15-25 keV) band based on the procedure outlined in Section~\ref{sec:spacial_dist}. \textbf{Middle-right:} The thermal-subtracted hard-band counts. \textbf{Right:} The SNR between the data, representing the local significance of the hard excess. 
    All images are smoothed by the binning process described in section \ref{sec:spacial_dist}, and the radius of the smoothing kernel is $r=10$~pixels.}
    \label{fig:thermsub}
\end{figure*}

\begin{figure*}
\begin{center}
\includegraphics[width=0.9\textwidth]{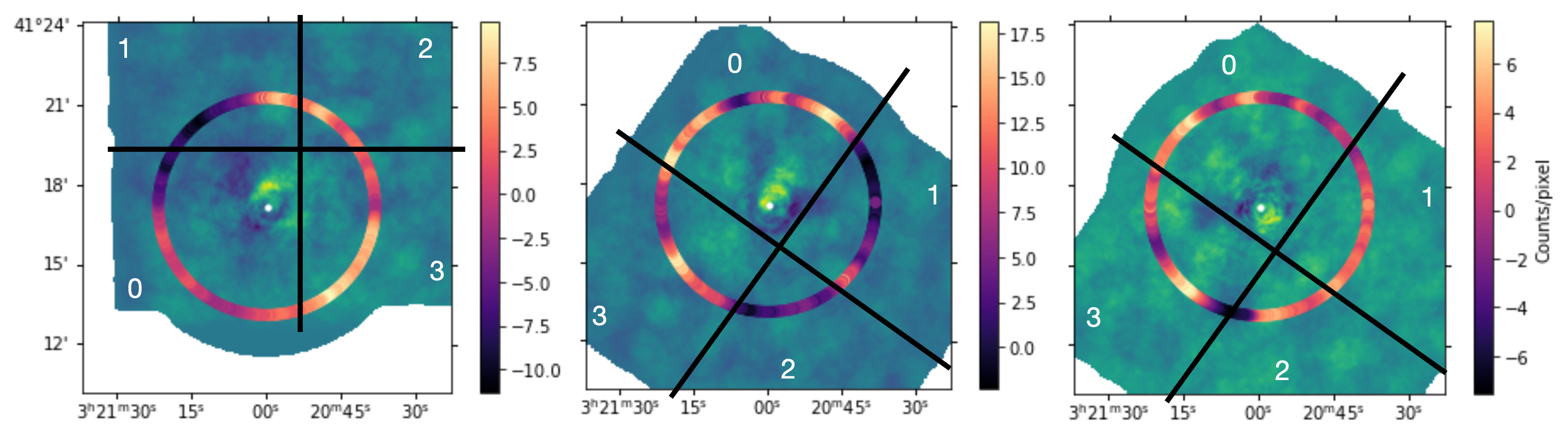}
\includegraphics[width=0.45\textwidth]{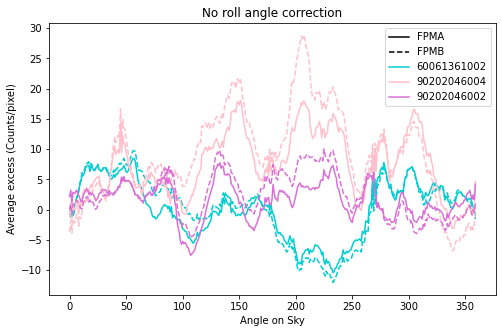}
\includegraphics[width=0.45\textwidth]{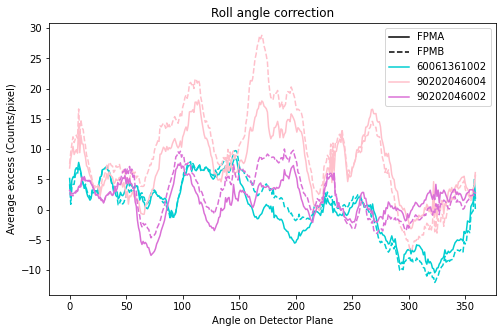}
 \end{center}
    \caption{\textbf{Top:} The excess in each observation (from left to right: 60061361002, 90202046004, and 90202046002) in units of counts, where the rounded extrusions along the edge of the observations are artifacts of the smoothing function. 
    The locations of the four detector chips are divided by the black cross-hatch and labeled with white numbers. 
    The color of the circle surrounding the AGN gives the summed excess in counts/pixel within a sector centered on that angular bin. 
    If the excesses originated from the cluster emission, it should be correlated between observations.
    \textbf{Bottom:} The excess as a function of angle for all three observations, either with respect to the sky (left) or the detector plane (right). The excess with respect to angle on the sky shows clear discrepancies between observations, as it also does with respect to roll angle.}
    \label{fig:excess_var}
\end{figure*}

\subsection{Spatial distribution of the hard excess}
\label{sec:spacial_dist}

\begin{figure}
    \centering
    \includegraphics[width=0.38\textwidth]{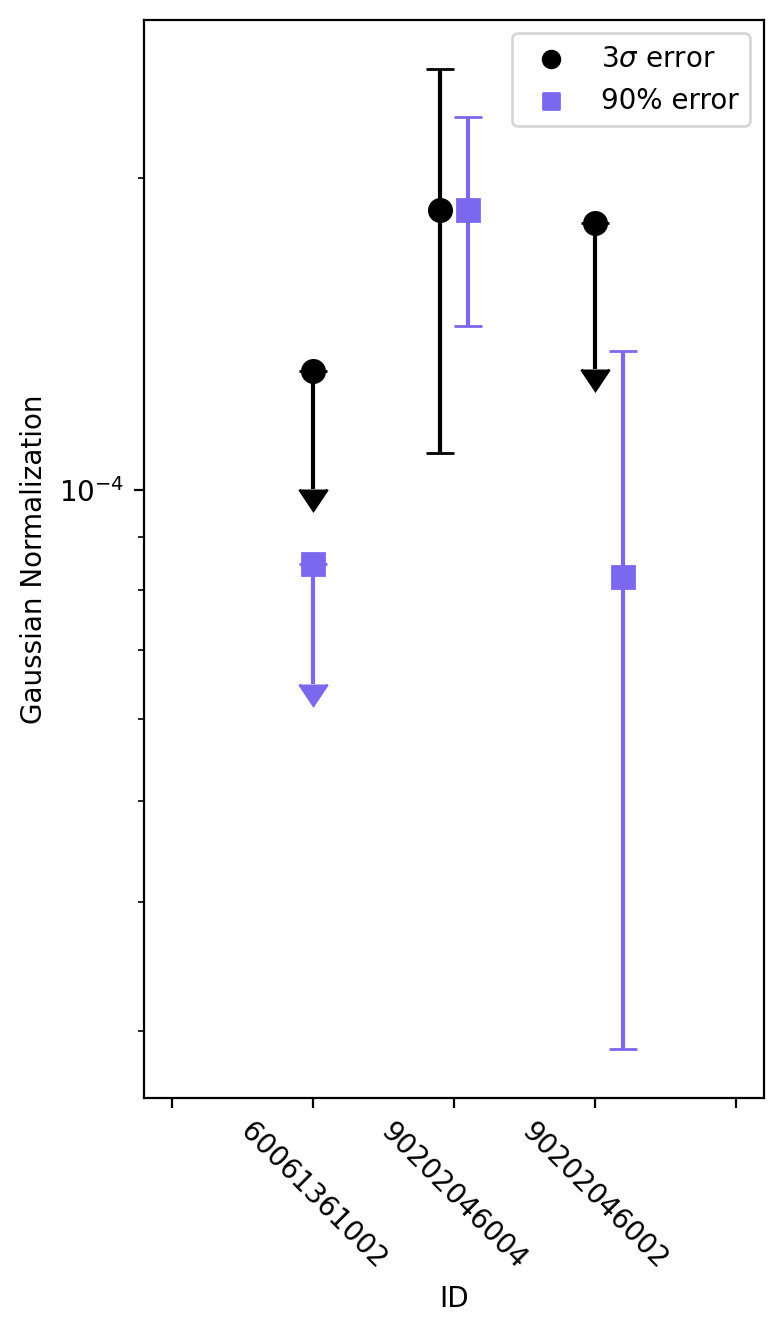}
    \caption{The normalization of the Gaussian component, which is defined as the total photons~cm$^{-2}$s$^{-1}$ contained within the line. This component represents the hard excess in the spectra, with $90 \%$ (purple) and $3 \sigma$ (black) error bars. 
    Where the lower bound dropped to zero, the normalization is represented as an upper limit.}
    \label{fig:ObsTest}
\end{figure}

In order to characterize the spatial distribution of the hard excess, we performed the following analysis for each \textit{NuSTAR} pixel within a box with the range $50.107\degree < \alpha < 49.796\degree$, $41.395\degree < \delta < 41.628\degree$, which contains the entire circular global region shown in Figure \ref{fig:glob}.

First, we centered a circular region ($r=10$~pixels) around the pixel. 
Using the soft-band (3-10 keV) image (background and AGN subtracted), we calculated the total number of soft counts within the region. 
We repeated this process for the hard-band (15-25 keV) image. 
Meanwhile, we placed the circle region in the \textit{NuSTAR} temperature map (Figure \ref{fig:tmap}) and took the mean pixel value as the average temperature within that region.

For each pixel, we then used the \textit{NuSTAR} response files (for both \textit{FPMA} and \textit{FPMB}) and the nearest ARFs to that pixel for each observation (found in the \textit{nucrossarf} libraries) to simulate an \textsc{APEC} model with $kT$ set to the the average temperature found within the $r=10$~pixel region.

From the \textsc{APEC} model, we extracted the flux count ratio between the soft (3-10 keV) and hard (15-25 keV, where the excess is strongest) energy bands. 
If all the hard-band counts from the ICM are due to thermal emission, this modeled soft-to-hard ratio will be equal to the \textit{actual} soft-to-hard count ratio of the data. 
So, by dividing the soft-band counts in the data by the modeled soft-to-hard ratio, we predicted the thermal counts in the hard (15-25 keV) band. 
We then subtracted this number from the actual hard-band data to obtain a map of the hard excess. These steps are summarized by the images in Figure \ref{fig:thermsub}.

By performing the above process for each pixel, we obtained an image of the hard excess and calculated its signal-to-noise ratio using the following equation:

\begin{center}
    \begin{equation} \label{eq:SNR}
    \frac{N_\text{total} - N_\text{thermal-modeled}}{\sqrt{N_\text{total}}}
    \end{equation}
\end{center}

Using this procedure,
we also mapped the spatial distribution of the hard excess in each individual observation.
In Figure~\ref{fig:excess_var}, we show the discrepancy between these three observations, which have roll angles of 89.7$^{\degree}$ (90202046002 and 90202046004) and $323.5^{\degree}$ (60061361002). 
The spatial distribution of the excess varies considerably amongst the pointings, and correcting for the roll angle does not eliminate the discrepancy significantly. 

In the light of this variation, we then fit the spectra from each individual observation to a 1T + Gauss model in order to check for the presence and significance of the excess on an observation-by-observation basis.
As shown in Figure \ref{fig:ObsTest}, the excess is only significant at the 3$\sigma$ level in observation 90202046004, and is not detected at even the 90\% level in observation 60061361002.

\section{Discussion}\label{sec:disc}
\subsection{Spectral modeling}

As described in section \ref{subsec:models}, we fit several models to our data in order to unveil the physical nature of the ICM in the Perseus cluster.
The statistics and best-fit parameters of these models are summarized in Table \ref{tab:pars}.

Fitting the ICM spectra to a single-temperature \textsc{APEC} (1T) yields a global temperature of 5.45$^{+0.04}_{-0.03}$~keV. 
This is consistent with the temperature found using Chandra \citep[$\sim$5.5~keV at $120$~kpc. See Figure \ref{fig:tmap} and][]{Pers_Chand_Sanders07,Schmidt20_Pers_Chandra_TempMap}. 
This 1T model describes the data well, but when additional components are added to the spectra, their best-fit parameters reveal a hard excess at $\sim 15 - 25$~keV.

When a second \textsc{APEC} component is added to describe the multi-temperature nature of the ICM, the best fit second temperature reaches nonphysically high values and hits the default ceiling for the parameter ($64$~keV).
At these high temperatures, the turn-over for bremsstrahlung emission occurs outside of the analyzed energy band, causing the second \textsc{APEC} component to mimic a powerlaw.
We conclude that the hard excess can not be described by thermal emission.

When we included a powerlaw component to account for possible non-thermal contribution to the ICM spectra (1T + IC), the best-fit model was not consistent with IC.
From the radio synchrotron index found by \citet{GM21}, the IC photon index should be $\Gamma = 2.2$.
But the best-fit powerlaw is extremely flat ($\Gamma = 0.254$), and fixing the powerlaw index to the expected value (2.2) was statistically disfavored over a powerlaw with a free $\Gamma$.
This implies that the hard excess is not IC emission, which is consistent with expectations.
The Perseus Cluster has a radio mini-halo \citep[discovered by][]{Burns92_minihalo_disc}, 
implying that the relativistic electron population is diffuse in radio mini-halos.
The radio synchrotron flux of these objects is notoriously faint,
so the IC signal from these relativistic electrons should also be relatively weak.
With a synchrotron flux density of 12.64~Jy \citep[230--470 MHz, see][]{GM17}, the Perseus Cluster's mini-halo is brighter than most by 3-4 orders of magnitude. 
However, if the global magnetic field is $\sim 50 \mu G$, as estimated by \cite{Perseus_Bcenter_Taylor06}, the IC flux would be $3 \times 10^{-16}$~erg~s$^{-1}$~cm$^{-2}$, which is five orders of magnitude fainter than the best-fit IC model (where $\Gamma$ is fixed).
So, although the hard excess is not IC, that finding is consistent with our expectations.

We fit the hard excess to three additional models that could describe a hard excess: a PyKappa model, which is the Non-Maxwellian equivalent of the \textsc{APEC}, a Comptonization model (expressed by a \textsc{SIMPL} convolution), and a Gaussian hump. 
Of these three, only the Gaussian was favored over the 1T, 2T, and 1T+IC models. 

Originally, the Gaussian model was motivated by the Compton Hump that occurs in obscured AGN spectra, but \citet{Rani18} find no evidence of reflection components (including the Compton Hump) in the AGN of the Perseus Cluster, so we did not expect to find a significant improvement. 
As expected, applying the Gaussian to the model that describes the AGN is only slightly favored over the 1T model presented in Table \ref{tab:pars}, where the AGN is described only by a powerlaw. 
However, adding the Gaussian to the model describing the ICM made a significant improvement. 
Although this component does not describe a known physical phenomenon, it provides useful information about the nature of the excess. 
Unlike a 1T+IC or 2T model, the Gaussian centered at hard energies has negligible contribution to soft-band emission.
Similarly, the best-fit flat powerlaw and 64~keV 2T models are trying to fit the excess while minimizing their effect on the soft-band spectrum.
This is demonstrated in Figure \ref{fig:model_comp}, which compares the two models that best describe the excess (1T~+~Gaussian and 1T~+~IC with a free $\Gamma$) with the three main physically motivated models (1T, 2T, and 1T~+~IC with $\Gamma$ fixed).
The components that attempt to describe the excess converge near $\sim17$~keV, demonstrating that all four of these models are fitting to the same feature in the hard end of the spectrum. 
These results suggest that the source of the hard excess only makes meaningful contributions to the emission around 15--20~keV, while the lower-energy spectrum is well described by pure thermal emission.

The wide Gaussian that best fits our data resembles an unaccounted-for effect in the background.
For example, a more recent analysis of archival data has revealed emission lines in the instrumental background that are not included in the current background model \citep[][]{Roach23_bgd_lines,Steve_Dissertation}{}{}.
Two of these emission lines are described by Lorentzians with $\mu = 13.07,~11.40$~keV and $\sigma = 20.00,~19.37$~keV. 
Neither of these lines align perfectly with the best-fit Gaussian for the hard excess---their locations are softer and their $\sigma$ values are wider than our Gaussian component---but fitting the excess to the harder of these two Lorentitzans (C-stat/d.o.f = 6494.68/5982) is statistically favored over the 1T and 1T + IC (fixed $\Gamma$) models. It is disfavored over the 2T, 1T + IC (free $\Gamma$), and 1T + Gauss models, so we cannot conclude that this new background feature explains the excess.

In addition to the Perseus cluster, we find evidence of features similar to the hard excess in other relaxed clusters, although the significance of the excess varies. 
Investigating these additional clusters is beyond the scope of this paper, but for future work, investigations of the hard excess would benefit from future X-ray telescopes with broader energy capabilities and lower background, such as the proposed High Energy X-ray Probe \citep[HEX-P][]{HEXP19}; reliable data above $25$~keV would allow for better characterization of the hard excess and test if it is truly represented by a Gaussian. 
Additionally, the low background of HEX-P would better constrain the nature of the excess and whether it is an effect of the sky, an intrinsic property of the ICM, or an instrumental effect of \textit{NuSTAR}.

\subsection{Implications of cross-ARF and background systematics}

In Figure \ref{fig:sys_2apec}, we show the results of applying the systematic errors in the \textsc{nucrossarf} and the \textit{NuSTAR} background models to our physically motivated models of the ICM (1T, 2T, and 1T + IC with both fixed and freed $\Gamma$). 
In this analysis, we generated 1000 iterations of the background models and AGN cross-ARFs, varying their contribution to the spectra within their systematic errors.
The right-most plot in the figure shows the difference in the C statistic ($\Delta$~C-stat) between a 2T and a given model (for $\Delta$~C~$>0$, the 2T model is favored).
This plot demonstrates that the 2T model is consistently favored over a 1T or physically realistic IC model ($\Gamma$ fixed at 2.2), while it did worse than a powerlaw with a free $\Gamma$.
Additionally, the second temperature in the 2T model hit the 64~keV ceiling in almost every case.

As shown in the middle panel of Figure~\ref{fig:sys_2apec}, $\Delta$~C-stat does not correlate with the strength of the AGN's contribution (within the 3.4\% systematic error). 
However, there is an obvious correlation with the aperture stray light component of the background ($S_d$). 
Increasing the value of the $S_d$ component caused our models to converge to a similar C-stat, while decreasing the contribution of $S_d$ widened the disparity between models and increased the significance of the hard excess. 
At the same time, increasing $S_d$ beyond its nominal value worsened the fit for \textit{all} models, which is strong evidence that increasing the $S_d$ does not describe the actual background.

We conclude that the hard excess cannot be explained by systematic uncertainties in either the background or AGN contribution, and its characteristics remain consistent regardless of changes to the background or AGN contribution.

\subsection{Variation of the hard excess}

Figure \ref{fig:thermsub} shows the AGN and background-subtracted image of excess photons in the $15-25$~keV energy band; these photons are not well described by thermal emission, and we take them to be responsible for the hard excess. 
There are regions of excess photons with signal-to-noise $>3$, which suggests that the hard excess may be localized.
However, the spatial distribution of the excess photons differ across observations, as shown in Figure \ref{fig:excess_var}. 
These variations occur on $kpc$ scales; if the excess were an attribute of the ICM, it should remain static over the time these observations were taken.
If the excess is due to an unaccounted-for detector effect, it may be correlated with the location of the detector chips. 
However, the roll angle correction shown in the bottom-right plot of Figure \ref{fig:excess_var} does not significantly improve the discrepancy between observations. 

In Figure \ref{fig:ObsTest}, we show that the strength of the excess (when represented by a Gaussian) is inconsistent between observations at the 90\% level. 
It is undetected at the 90\% level in observation 60061361002, which has the lowest exposure time of the three observations, and is only detected at the 3$\sigma$ level in observation 90202046004, which has the highest exposure time.
To investigate the chance that the hard excess is present---simply not detectable---in observation 60061361002, we present the signal-to-noise ratio of the excess (SNR$_\text{excess}$~=~$N_{\text{excess}}/ \sqrt{N_{\text{total}}}$, where $N_{\text{excess}}$ is the modeled number of photons in the Gaussian model component and $N_\text{total}$ is the total number of events present, both in the 15-25 keV band) of our observations in Table \ref{tab:SNR}.
While SNR appears to increase with exposure time, the difference in SNR$_\text{excess}$ between observations 90202046002 and 90202046004 far exceeds what we would expect given the $0.3$~ks difference in exposure times.
While this does not completely rule out the possibility that the excess is a physical property of the ICM, it is more in line with expectations for a transient background feature or other instrumental effect.

\begin{table*}
    \centering
    \caption{Signal to Noise Ratio of Excess\label{tab:SNR}}
    \begin{tabular}{lcccccc}
     \hline 
     \hline
     \\[-0.75em]
          & Cleaned Exposure & Rate (excess) &  & Rate (total) & &  \\
         Obs. ID & (ks) & (photons/s) & Photons (excess) & (photons/s) & Photons (total) & SNR$_\text{excess}$\\\\[-0.75em]
         \hline
         \\[-0.5em]
         
         90202046002 & 27.5 ks & 0.011 & 309.4 & 0.122 & 3445.8 & 5.27
         \\
         90202046004 & 28.2 ks & 0.018 & 507.3 & 0.132 & 3736.5 & 8.30
         \\
         60061361002 & 14.6 ks & 0.007 & 107.3 & 0.125 & 1774.6 & 2.55
         \\
      \hline
    \end{tabular}
\end{table*}

\subsection{Thermal distribution}
Figure \ref{fig:tmap} shows the \textit{NuSTAR} and \textit{Chandra} temperature map of the ICM of the Perseus Cluster.  
The comparison shows that the temperature structure is consistent between both telescopes, with a cool core and hot ($\sim 9$~keV) gas outside the core. 

The main discrepancy between the maps is that \textit{NuSTAR} measures higher temperatures across the cluster. 
It is a long-standing issue that different instruments---including \textit{NuSTAR} and \textit{Chandra}---measure different temperatures when observing the same ICM.
The calibration differences between these telescopes are not well understood, though work is being done to quantify them \citep{Potter23}.
In the simplest form, the discrepancy that we see between the measurements in Figure \ref{fig:tmap} is consistent with our expectations; \textit{NuSTAR's} harder bandpass biases it towards measuring higher temperatures when a multi-temperature gas is present in a spectrum being fit to a single \texttt{apec} model. 

Both \textit{NuSTAR} and \textit{Chandra} clearly observe the Perseus Cluster's cool core, and the cold front---the sharp discontinuity between the cool core and the hot outer ICM \citep[first discovered by][]{Fabian11}---is apparent in both temperature maps.

\subsection{Magnetic field calculation} \label{subsec:B_calc}

In Section~\ref{sec:Non-Thermal}, we obtained an upper limit to the IC flux of $F_{IC(4-25~\text{keV})} \leq 1.5 \times 10^{-11}$~erg~s$^{-1}$~cm$^{-2}$. 
From this, we can constrain the magnetic field using the following equation as derived by \cite{B_review_04}:
\begin{multline}
    (B [ \mu G])^{1 + \alpha} =
    h(\alpha)
    \frac{F_{syn(\nu_r)}}{F_{IC(E_1 - E_2)}}
    (1 + z)^{3 + \alpha}
    (0.0545 \nu_r )^\alpha \\
    \times (E_2^{1-\alpha} - E_1^{1-\alpha})\, ,
\label{eq:B_IC}
\end{multline}
where the energy range of $F_{IC(E_1-E_2)}$ is 4-25~keV, $\alpha$ is the spectral index of the radio synchrotron emission, $F_{sync(\nu_r)}$ is the synchrotron flux at frequency $\nu_r$, and $h(\alpha)$ is tabulated in Table 2 of \cite{B_review_04}.

\cite{GM21} report that the synchrotron flux of the radio mini-halo is 12.64~Jy (230–470~MHz; for the following calculation, we use the midpoint frequency of $\nu_r=350$~MHz) and find the sychrotron index to be $\alpha = 1.2$, which gives $h(\alpha)=-1.37 \times 10^{-14}$.
Folding these values into Equation \ref{eq:B_IC}, we constrain the magnetic field to be $\geq 0.35~\mu G$.
Although this measurement is consistent with previous estimates of the magnetic field in the Perseus cluster, it is an order of magnitude lower than the lower limit found by \citet{Perseus_B_Aleksi12} and two orders of magnitude lower than the the estimate by \citet{Perseus_Bcenter_Taylor06}. 
Our lower limit is consistent both with these estimates and with the expectation that the magnetic field in the cool cores of clusters should be stronger than the globally averaged value of $B$; {\it NuSTAR} measurements of other clusters yield similar lower limits \citep{RB23}, so a detection of IC emission in the core of Perseus at this level would be unexpected.

\section{Conclusions}\label{sec:conc}
\begin{itemize}
    \item We studied the thermal structure of the Perseus cluster using three \textit{NuSTAR} observations, and the results are consistent (within known systematic effects) with \textit{Chandra} temperature measurements, as shown in Figure \ref{fig:tmap}.
    
    \item We find evidence of a hard excess in two of the three observations that cannot be well modeled by thermal emission or explained by a known systematic error in the background or AGN cross-ARF modeling. The excess is best described by a wide Gaussian at 17~keV, and it is not consistent with a realistic inverse Compton model. 

    \item The strength and spatial distribution of the hard excess is inconsistent across observations, which possibly points to a non-astrophysical origin.
    
    \item By applying a physically motivated IC model to the excess, we constrain the cluster's global magnetic field to be $\gtrsim 0.35~\mu G$, which is 1--2 orders of magnitude lower than (and consistent with) previous measurements made by \cite{Perseus_B_Aleksi12} and \cite{Perseus_Bcenter_Taylor06}.

    \item We report that \textsc{nucrossarf} models scattered light from point sources with an accuracy of 3.4\%, which we report as its systematic uncertainty (see Appendix~\ref{CR systematics} for details).

\end{itemize}

\section*{Acknowledgments}
We thank the anonymous referee for useful comments that improved the work. This work was made possible with funding from the NASA Astrophysics Data Analysis Program grant 80NSSC20K1000.
This work made use of data from the NuSTAR mission, which is led by the California Institute of Technology, managed by the Jet Propulsion Laboratory, and funded by the National Aeronautics and Space Administration.
Additionally, this research has made use of data and software provided by the High Energy Astrophysics Science Archive Research Center (HEASARC), which is a service of the Astrophysics Science Division at NASA/GSFC and the High Energy Astrophysics Division of the Smithsonian Astrophysical Observatory.
This work also made use of Astropy---a community-developed core Python package for Astronomy (Astropy Collaboration, 2013)---, numpy---Van der Walt and Colbert 2011, Computing in Science \& Engineering 13, 22---, and scipy---a community-developed open source software for scientific computing in Python \citep{astropy,SciPy,numpy}.
Finally, S.C. and D.R.W would like to thank Kristin Madsen for her valuable discussions at the HEAD2023 meeting.

\bibliography{main}{}

\begin{thebibliography}{}
\expandafter\ifx\csname natexlab\endcsname\relax\def\natexlab#1{#1}\fi
\providecommand{\url}[1]{\href{#1}{#1}}
\providecommand{\dodoi}[1]{doi:~\href{http://doi.org/#1}{\nolinkurl{#1}}}
\providecommand{\doeprint}[1]{\href{http://ascl.net/#1}{\nolinkurl{http://ascl.net/#1}}}
\providecommand{\doarXiv}[1]{\href{https://arxiv.org/abs/#1}{\nolinkurl{https://arxiv.org/abs/#1}}}

\bibitem[{{Akahori} \& {Yoshikawa}(2010)}]{NonMax_GC_Mergers_Akahori10}
{Akahori}, T., \& {Yoshikawa}, K. 2010, \pasj, 62, 335,
  \dodoi{10.1093/pasj/62.2.335}

\bibitem[{{Aleksi{\'c}} {et~al.}(2012){Aleksi{\'c}}, {Alvarez}, {Antonelli},
  {Antoranz}, {Asensio}, {Backes}, {Barres de Almeida}, {Barrio}, {Bastieri},
  {Becerra Gonz{\'a}lez}, {Bednarek}, {Berdyugin}, {Berger}, {Bernardini},
  {Biland}, {Blanch}, {Bock}, {Boller}, {Bonnoli}, {Borla Tridon}, {Braun},
  {Bretz}, {Ca{\~n}ellas}, {Carmona}, {Carosi}, {Colin}, {Colombo},
  {Contreras}, {Cortina}, {Cossio}, {Covino}, {Dazzi}, {de Angelis}, {de
  Caneva}, {de Cea Del Pozo}, {de Lotto}, {Delgado Mendez}, {Diago Ortega},
  {Doert}, {Dom{\'\i}nguez}, {Dominis Prester}, {Dorner}, {Doro}, {Eisenacher},
  {Elsaesser}, {Ferenc}, {Fonseca}, {Font}, {Fruck}, {Garc{\'\i}a L{\'o}pez},
  {Garczarczyk}, {Garrido}, {Giavitto}, {Godinovi{\'c}}, {Gozzini}, {Hadasch},
  {H{\"a}fner}, {Herrero}, {Hildebrand}, {H{\"o}hne-M{\"o}nch}, {Hose},
  {Hrupec}, {Jogler}, {Kellermann}, {Klepser}, {Kr{\"a}henb{\"u}hl}, {Krause},
  {Kushida}, {La Barbera}, {Lelas}, {Leonardo}, {Lewandowska}, {Lindfors},
  {Lombardi}, {L{\'o}pez}, {L{\'o}pez}, {L{\'o}pez-Oramas}, {Lorenz},
  {Makariev}, {Maneva}, {Mankuzhiyil}, {Mannheim}, {Maraschi}, {Mariotti},
  {Mart{\'\i}nez}, {Mazin}, {Meucci}, {Miranda}, {Mirzoyan}, {Mold{\'o}n},
  {Moralejo}, {Munar-Adrover}, {Niedzwiecki}, {Nieto}, {Nilsson}, {Nowak},
  {Orito}, {Paiano}, {Paneque}, {Paoletti}, {Pardo}, {Paredes}, {Partini},
  {Perez-Torres}, {Persic}, {Peruzzo}, {Pilia}, {Pochon}, {Prada}, {Prada
  Moroni}, {Prandini}, {Puerto Gimenez}, {Puljak}, {Reichardt}, {Reinthal},
  {Rhode}, {Rib{\'o}}, {Rico}, {R{\"u}gamer}, {Saggion}, {Saito}, {Saito},
  {Salvati}, {Satalecka}, {Scalzotto}, {Scapin}, {Schultz}, {Schweizer},
  {Shayduk}, {Shore}, {Sillanp{\"a}{\"a}}, {Sitarek}, {Snidaric}, {Sobczynska},
  {Spanier}, {Spiro}, {Stamatescu}, {Stamerra}, {Steinke}, {Storz}, {Strah},
  {Sun}, {Suri{\'c}}, {Takalo}, {Takami}, {Tavecchio}, {Temnikov},
  {Terzi{\'c}}, {Tescaro}, {Teshima}, {Tibolla}, {Torres}, {Treves},
  {Uellenbeck}, {Vankov}, {Vogler}, {Wagner}, {Weitzel}, {Zabalza}, {Zandanel},
  {Zanin}, {MAGIC Collaboration}, {Pfrommer}, \& {Pinzke}}]{Perseus_B_Aleksi12}
{Aleksi{\'c}}, J., {Alvarez}, E.~A., {Antonelli}, L.~A., {et~al.} 2012, \aap,
  541, A99, \dodoi{10.1051/0004-6361/201118502}

\bibitem[{{Ballarati} {et~al.}(1981){Ballarati}, {Feretti}, {Ficarra},
  {Giovannini}, {Nanni}, {Olori}, \& {Gavazzi}}]{Relic_class81}
{Ballarati}, B., {Feretti}, L., {Ficarra}, A., {et~al.} 1981, \aap, 100, 323

\bibitem[{{Balokovi{\'c}} {et~al.}(2016){Balokovi{\'c}}, {Paneque}, {Madejski},
  {Furniss}, {Chiang}, {Ajello}, {Alexander}, {Barret}, {Blandford}, {Boggs},
  {Christensen}, {Craig}, {Forster}, {Giommi}, {Grefenstette}, {Hailey},
  {Harrison}, {Hornstrup}, {Kitaguchi}, {Koglin}, {Madsen}, {Mao}, {Miyasaka},
  {Mori}, {Perri}, {Pivovaroff}, {Puccetti}, {Rana}, {Stern}, {Tagliaferri},
  {Urry}, {Westergaard}, {Zhang}, {Zoglauer}, {NuSTAR Team}, {Archambault},
  {Archer}, {Barnacka}, {Benbow}, {Bird}, {Buckley}, {Bugaev}, {Cerruti},
  {Chen}, {Ciupik}, {Connolly}, {Cui}, {Dickinson}, {Dumm}, {Eisch}, {Falcone},
  {Feng}, {Finley}, {Fleischhack}, {Fortson}, {Griffin}, {Griffiths}, {Grube},
  {Gyuk}, {Huetten}, {H{\r{a}}kansson}, {Holder}, {Humensky}, {Johnson},
  {Kaaret}, {Kertzman}, {Khassen}, {Kieda}, {Krause}, {Krennrich}, {Lang},
  {Maier}, {McArthur}, {Meagher}, {Moriarty}, {Nelson}, {Nieto}, {Ong}, {Park},
  {Pohl}, {Popkow}, {Pueschel}, {Reynolds}, {Richards}, {Roache}, {Santander},
  {Sembroski}, {Shahinyan}, {Smith}, {Staszak}, {Telezhinsky}, {Todd}, {Tucci},
  {Tyler}, {Vincent}, {Weinstein}, {Wilhelm}, {Williams}, {Zitzer}, {VERITAS
  Collaboration}, {Ahnen}, {Ansoldi}, {Antonelli}, {Antoranz}, {Babic},
  {Banerjee}, {Bangale}, {Barres de Almeida}, {Barrio}, {Becerra Gonz{\'a}lez},
  {Bednarek}, {Bernardini}, {Biasuzzi}, {Biland}, {Blanch}, {Bonnefoy},
  {Bonnoli}, {Borracci}, {Bretz}, {Carmona}, {Carosi}, {Chatterjee}, {Clavero},
  {Colin}, {Colombo}, {Contreras}, {Cortina}, {Covino}, {Da Vela}, {Dazzi}, {De
  Angelis}, {De Lotto}, {de O{\~n}a Wilhelmi}, {Delgado Mendez}, {Di Pierro},
  {Dominis Prester}, {Dorner}, {Doro}, {Einecke}, {Elsaesser},
  {Fern{\'a}ndez-Barral}, {Fidalgo}, {Fonseca}, {Font}, {Frantzen}, {Fruck},
  {Galindo}, {Garc{\'\i}a L{\'o}pez}, {Garczarczyk}, {Garrido Terrats}, {Gaug},
  {Giammaria}, {Glawion (Eisenacher}, {Godinovi{\'c}}, {Gonz{\'a}lez
  Mu{\~n}oz}, {Guberman}, {Hahn}, {Hanabata}, {Hayashida}, {Herrera}, {Hose},
  {Hrupec}, {Hughes}, {Idec}, {Kodani}, {Konno}, {Kubo}, {Kushida}, {La
  Barbera}, {Lelas}, {Lindfors}, {Lombardi}, {Longo}, {L{\'o}pez},
  {L{\'o}pez-Coto}, {L{\'o}pez-Oramas}, {Lorenz}, {Majumdar}, {Makariev},
  {Mallot}, {Maneva}, {Manganaro}, {Mannheim}, {Maraschi}, {Marcote},
  {Mariotti}, {Mart{\'\i}nez}, {Mazin}, {Menzel}, {Miranda}, {Mirzoyan},
  {Moralejo}, {Moretti}, {Nakajima}, {Neustroev}, {Niedzwiecki}, {Nievas
  Rosillo}, {Nilsson}, {Nishijima}, {Noda}, {Orito}, {Overkemping}, {Paiano},
  {Palacio}, {Palatiello}, {Paoletti}, {Paredes}, {Paredes-Fortuny}, {Persic},
  {Poutanen}, {Prada Moroni}, {Prandini}, {Puljak}, {Rhode}, {Rib{\'o}},
  {Rico}, {Rodriguez Garcia}, {Saito}, {Satalecka}, {Scapin}, {Schultz},
  {Schweizer}, {Shore}, {Sillanp{\"a}{\"a}}, {Sitarek}, {Snidaric},
  {Sobczynska}, {Stamerra}, {Steinbring}, {Strzys}, {Takalo}, {Takami},
  {Tavecchio}, {Temnikov}, {Terzi{\'c}}, {Tescaro}, {Teshima}, {Thaele},
  {Torres}, {Toyama}, {Treves}, {Verguilov}, {Vovk}, {Ward}, {Will}, {Wu},
  {Zanin}, {MAGIC Collaboration}, {Perkins}, {Verrecchia}, {Leto},
  {B{\"o}ttcher}, {Villata}, {Raiteri}, {Acosta-Pulido}, {Bachev}, {Berdyugin},
  {Blinov}, {Carnerero}, {Chen}, {Chinchilla}, {Damljanovic}, {Eswaraiah},
  {Grishina}, {Ibryamov}, {Jordan}, {Jorstad}, {Joshi}, {Kopatskaya},
  {Kurtanidze}, {Kurtanidze}, {Larionova}, {Larionova}, {Larionov}, {Latev},
  {Lin}, {Marscher}, {Mokrushina}, {Morozova}, {Nikolashvili}, {Semkov},
  {Smith}, {Strigachev}, {Troitskaya}, {Troitsky}, {Vince}, {Barnes},
  {G{\"u}ver}, {Moody}, {Sadun}, {Sun}, {Hovatta}, {Richards}, {Max-Moerbeck},
  {Readhead}, {L{\"a}hteenm{\"a}ki}, {Tornikoski}, {Tammi}, {Ramakrishnan},
  {Reinthal}, {Angelakis}, {Fuhrmann}, {Myserlis}, {Karamanavis}, {Sievers},
  {Ungerechts}, \& {Zensus}}]{MRK421}
{Balokovi{\'c}}, M., {Paneque}, D., {Madejski}, G., {et~al.} 2016, \apj, 819,
  156, \dodoi{10.3847/0004-637X/819/2/156}

\bibitem[{{Brunetti} {et~al.}(2001){Brunetti}, {Setti}, {Feretti}, \&
  {Giovannini}}]{B_coma_Brunetti01}
{Brunetti}, G., {Setti}, G., {Feretti}, L., \& {Giovannini}, G. 2001, \mnras,
  320, 365, \dodoi{10.1046/j.1365-8711.2001.03978.x}

\bibitem[{{Burns} {et~al.}(1992){Burns}, {Sulkanen}, {Gisler}, \&
  {Perley}}]{Burns92_minihalo_disc}
{Burns}, J.~O., {Sulkanen}, M.~E., {Gisler}, G.~R., \& {Perley}, R.~A. 1992,
  \apjl, 388, L49, \dodoi{10.1086/186327}

\bibitem[{{Cash}(1979)}]{cash79}
{Cash}, W. 1979, \apj, 228, 939, \dodoi{10.1086/156922}

\bibitem[{{Cui} {et~al.}(2019){Cui}, {Foster}, {Yuasa}, \&
  {Smith}}]{Cui19_kappa}
{Cui}, X., {Foster}, A.~R., {Yuasa}, T., \& {Smith}, R.~K. 2019, \apj, 887,
  182, \dodoi{10.3847/1538-4357/ab5304}

\bibitem[{{En{\ss}lin} \& {Br{\"u}ggen}(2002)}]{Relic_formation02}
{En{\ss}lin}, T.~A., \& {Br{\"u}ggen}, M. 2002, \mnras, 331, 1011,
  \dodoi{10.1046/j.1365-8711.2002.05261.x}

\bibitem[{Fabian {et~al.}(2006)Fabian, Sanders, Taylor, Allen, Crawford,
  Johnstone, \& Iwasawa}]{Fabian06}
Fabian, A.~C., Sanders, J.~S., Taylor, G.~B., {et~al.} 2006, Monthly Notices of
  the Royal Astronomical Society, 366, 417,
  \dodoi{10.1111/j.1365-2966.2005.09896.x}

\bibitem[{Fabian {et~al.}(2015)Fabian, Walker, Pinto, Russell, \&
  Edge}]{Fabian15}
Fabian, A.~C., Walker, S.~A., Pinto, C., Russell, H.~R., \& Edge, A.~C. 2015,
  Effects of the variability of the nucleus of NGC1275 on X-ray observations of
  the surrounding intracluster medium,  arXiv,
  \dodoi{10.48550/ARXIV.1505.03754}

\bibitem[{Fabian {et~al.}(2011)Fabian, Sanders, Allen, Canning, Churazov,
  Crawford, Forman, GaBany, Hlavacek-Larrondo, Johnstone, Russell, Reynolds,
  Salomé, Taylor, \& Young}]{Fabian11}
Fabian, A.~C., Sanders, J.~S., Allen, S.~W., {et~al.} 2011, Monthly Notices of
  the Royal Astronomical Society, 418, 2154–2164,
  \dodoi{10.1111/j.1365-2966.2011.19402.x}

\bibitem[{Gendron-Marsolais {et~al.}(2017)Gendron-Marsolais, Hlavacek-Larrondo,
  van Weeren, Clarke, Fabian, Intema, Taylor, Blundell, \& Sanders}]{GM17}
Gendron-Marsolais, M., Hlavacek-Larrondo, J., van Weeren, R.~J., {et~al.} 2017,
  Monthly Notices of the Royal Astronomical Society, 469, 3872,
  \dodoi{10.1093/mnras/stx1042}

\bibitem[{{Gendron-Marsolais} {et~al.}(2021){Gendron-Marsolais}, {Hull},
  {Perley}, {Rudnick}, {Kraft}, {Hlavacek-Larrondo}, {Fabian}, {Roediger}, {van
  Weeren}, {Richard-Laferri{\`e}re}, {Golden-Marx}, {Arakawa}, \&
  {McBride}}]{GM21}
{Gendron-Marsolais}, M.~L., {Hull}, C.~L.~H., {Perley}, R., {et~al.} 2021,
  \apj, 911, 56, \dodoi{10.3847/1538-4357/abddbb}

\bibitem[{{Giacintucci} {et~al.}(2014){Giacintucci}, {Markevitch}, {Brunetti},
  {ZuHone}, {Venturi}, {Mazzotta}, \&
  {Bourdin}}]{minihalo_fluxes_Giacintucci14}
{Giacintucci}, S., {Markevitch}, M., {Brunetti}, G., {et~al.} 2014, \apj, 795,
  73, \dodoi{10.1088/0004-637X/795/1/73}

\bibitem[{{Gitti} {et~al.}(2002){Gitti}, {Brunetti}, \&
  {Setti}}]{minihalo_turbulence_model_Perseus_Gitti02}
{Gitti}, M., {Brunetti}, G., \& {Setti}, G. 2002, \aap, 386, 456,
  \dodoi{10.1051/0004-6361:20020284}

\bibitem[{{Govoni} \& {Feretti}(2004)}]{B_review_04}
{Govoni}, F., \& {Feretti}, L. 2004, International Journal of Modern Physics D,
  13, 1549, \dodoi{10.1142/S0218271804005080}

\bibitem[{{Hahn} \& {Savin}(2015)}]{NonMax_Fe_Hahn15}
{Hahn}, M., \& {Savin}, D.~W. 2015, \apj, 800, 68,
  \dodoi{10.1088/0004-637X/800/1/68}

\bibitem[{Harris {et~al.}(2020)Harris, Millman, van~der Walt, Gommers,
  Virtanen, Cournapeau, Wieser, Taylor, Berg, Smith, Kern, Picus, Hoyer, van
  Kerkwijk, Brett, Haldane, del R{\'{i}}o, Wiebe, Peterson,
  G{\'{e}}rard-Marchant, Sheppard, Reddy, Weckesser, Abbasi, Gohlke, \&
  Oliphant}]{numpy}
Harris, C.~R., Millman, K.~J., van~der Walt, S.~J., {et~al.} 2020, Nature, 585,
  357, \dodoi{10.1038/s41586-020-2649-2}

\bibitem[{{Madsen} {et~al.}(2019){Madsen}, {Hickox}, {Bachetti}, {Stern},
  {Gellert}, {Garc{\'\i}a}, {Kara}, {Brandt}, {Krawczynski}, {Lohfink},
  {Brenneman}, {Christensen}, {Middleton}, {Hornstrup}, {Matt}, {Jaodand},
  {Lansbury}, {Ricci}, {Fuerst}, {Ballantyne}, {Walton}, {Fabian}, {Della
  Monica Ferreira}, {Pottschmidt}, {Miller}, {Windt}, {Balokovi{\'c}},
  {Kamraj}, {Wilms}, {Heida}, {Alexander}, {Boorman}, {Wik}, {Vogel},
  {Earnshaw}, {Descalle}, {Civano}, {Fornasini}, {Grindlay}, {Zhang},
  {Hornschemeier}, \& {Craig}}]{HEXP19}
{Madsen}, K., {Hickox}, R., {Bachetti}, M., {et~al.} 2019, in Bulletin of the
  American Astronomical Society, Vol.~51, 166

\bibitem[{{Mirakhor} {et~al.}(2022){Mirakhor}, {Walker}, {Runge}, \&
  {Diwanji}}]{Mirakhor22}
{Mirakhor}, M.~S., {Walker}, S.~A., {Runge}, J., \& {Diwanji}, P. 2022, arXiv
  e-prints, arXiv:2208.09553.
\newblock \doarXiv{2208.09553}

\bibitem[{{Mushotzky} {et~al.}(1978){Mushotzky}, {Serlemitsos}, {Smith},
  {Boldt}, \& {Holt}}]{Mushotzky78_Pers_Xlumin}
{Mushotzky}, R.~F., {Serlemitsos}, P.~J., {Smith}, B.~W., {Boldt}, E.~A., \&
  {Holt}, S.~S. 1978, \apj, 225, 21, \dodoi{10.1086/156465}

\bibitem[{{Nevalainen} {et~al.}(2004){Nevalainen}, {Oosterbroek}, {Bonamente},
  \& {Colafrancesco}}]{FirstComa_IC_04}
{Nevalainen}, J., {Oosterbroek}, T., {Bonamente}, M., \& {Colafrancesco}, S.
  2004, \apj, 608, 166, \dodoi{10.1086/381231}

\bibitem[{{Potter} {et~al.}(2023){Potter}, {T{\"u}mer}, {Wang}, {Wik},
  {Maughan}, \& {Schellenberger}}]{Potter23}
{Potter}, C., {T{\"u}mer}, A., {Wang}, Q. H.~S., {et~al.} 2023, arXiv e-prints,
  arXiv:2309.11743, \dodoi{10.48550/arXiv.2309.11743}

\bibitem[{{Price-Whelan} {et~al.}(2018){Price-Whelan}, {Sip{\H{o}}cz},
  {G{\"u}nther}, {Lim}, {Crawford}, {Conseil}, {Shupe}, {Craig}, {Dencheva},
  {Ginsburg}, {VanderPlas}, {Bradley}, {P{\'e}rez-Su{\'a}rez}, {de Val-Borro},
  {Paper Contributors}, {Aldcroft}, {Cruz}, {Robitaille}, {Tollerud},
  {Coordination Committee}, {Ardelean}, {Babej}, {Bach}, {Bachetti}, {Bakanov},
  {Bamford}, {Barentsen}, {Barmby}, {Baumbach}, {Berry}, {Biscani}, {Boquien},
  {Bostroem}, {Bouma}, {Brammer}, {Bray}, {Breytenbach}, {Buddelmeijer},
  {Burke}, {Calderone}, {Cano Rodr{\'\i}guez}, {Cara}, {Cardoso}, {Cheedella},
  {Copin}, {Corrales}, {Crichton}, {D{\textquoteright}Avella}, {Deil},
  {Depagne}, {Dietrich}, {Donath}, {Droettboom}, {Earl}, {Erben}, {Fabbro},
  {Ferreira}, {Finethy}, {Fox}, {Garrison}, {Gibbons}, {Goldstein}, {Gommers},
  {Greco}, {Greenfield}, {Groener}, {Grollier}, {Hagen}, {Hirst}, {Homeier},
  {Horton}, {Hosseinzadeh}, {Hu}, {Hunkeler}, {Ivezi{\'c}}, {Jain}, {Jenness},
  {Kanarek}, {Kendrew}, {Kern}, {Kerzendorf}, {Khvalko}, {King}, {Kirkby},
  {Kulkarni}, {Kumar}, {Lee}, {Lenz}, {Littlefair}, {Ma}, {Macleod},
  {Mastropietro}, {McCully}, {Montagnac}, {Morris}, {Mueller}, {Mumford},
  {Muna}, {Murphy}, {Nelson}, {Nguyen}, {Ninan}, {N{\"o}the}, {Ogaz}, {Oh},
  {Parejko}, {Parley}, {Pascual}, {Patil}, {Patil}, {Plunkett}, {Prochaska},
  {Rastogi}, {Reddy Janga}, {Sabater}, {Sakurikar}, {Seifert}, {Sherbert},
  {Sherwood-Taylor}, {Shih}, {Sick}, {Silbiger}, {Singanamalla}, {Singer},
  {Sladen}, {Sooley}, {Sornarajah}, {Streicher}, {Teuben}, {Thomas},
  {Tremblay}, {Turner}, {Terr{\'o}n}, {van Kerkwijk}, {de la Vega}, {Watkins},
  {Weaver}, {Whitmore}, {Woillez}, {Zabalza}, \& {Contributors}}]{astropy}
{Price-Whelan}, A.~M., {Sip{\H{o}}cz}, B.~M., {G{\"u}nther}, H.~M., {et~al.}
  2018, \aj, 156, 123, \dodoi{10.3847/1538-3881/aabc4f}

\bibitem[{{Rani} {et~al.}(2018){Rani}, {Madejski}, {Mushotzky}, {Reynolds}, \&
  {Hodgson}}]{Rani18}
{Rani}, B., {Madejski}, G.~M., {Mushotzky}, R.~F., {Reynolds}, C., \&
  {Hodgson}, J.~A. 2018, \apjl, 866, L13, \dodoi{10.3847/2041-8213/aae48f}

\bibitem[{{Roach} {et~al.}(2023){Roach}, {Rossland}, {Ng}, {Perez}, {Beacom},
  {Grefenstette}, {Horiuchi}, {Krivonos}, \& {Wik}}]{Roach23_bgd_lines}
{Roach}, B.~M., {Rossland}, S., {Ng}, K. C.~Y., {et~al.} 2023, \prd, 107,
  023009, \dodoi{10.1103/PhysRevD.107.023009}

\bibitem[{{Rojas Bolivar} {et~al.}(2023){Rojas Bolivar}, {Wik}, {T{\"u}mer},
  {Gastaldello}, {Hlavacek-Larrondo}, {Nulsen}, {Vacca}, {Madejski}, {Sun},
  {Sarazin}, {Sanders}, {Caprioli}, {Grefenstette}, \& {Westergaard}}]{RB23}
{Rojas Bolivar}, R., {Wik}, D., {T{\"u}mer}, A., {et~al.} 2023, arXiv e-prints,
  arXiv:2308.00969, \dodoi{10.48550/arXiv.2308.00969}

\bibitem[{{Rojas Bolivar} {et~al.}(2021){Rojas Bolivar}, {Wik}, {Giacintucci},
  {Gastaldello}, {Hornstrup}, {Westergaard}, \& {Madejski}}]{RB21}
{Rojas Bolivar}, R.~A., {Wik}, D.~R., {Giacintucci}, S., {et~al.} 2021, \apj,
  906, 87, \dodoi{10.3847/1538-4357/abcbf7}

\bibitem[{Rossland(2022)}]{Steve_Dissertation}
Rossland, S. 2022, Doctoral dissertation, University of Utah, Salt Lake City,
  UT

\bibitem[{{Sanders} \& {Fabian}(2007)}]{Pers_Chand_Sanders07}
{Sanders}, J.~S., \& {Fabian}, A.~C. 2007, \mnras, 381, 1381,
  \dodoi{10.1111/j.1365-2966.2007.12347.x}

\bibitem[{{Schmidt} {et~al.}(2002){Schmidt}, {Fabian}, \&
  {Sanders}}]{Schmidt20_Pers_Chandra_TempMap}
{Schmidt}, R.~W., {Fabian}, A.~C., \& {Sanders}, J.~S. 2002, \mnras, 337, 71,
  \dodoi{10.1046/j.1365-8711.2002.05804.x}

\bibitem[{{Simionescu} {et~al.}(2011){Simionescu}, {Allen}, {Mantz}, {Werner},
  {Takei}, {Morris}, {Fabian}, {Sanders}, {Nulsen}, {George}, \&
  {Taylor}}]{Simionescu11_Pers_Pars}
{Simionescu}, A., {Allen}, S.~W., {Mantz}, A., {et~al.} 2011, Science, 331,
  1576, \dodoi{10.1126/science.1200331}

\bibitem[{{Smith} {et~al.}(2001){Smith}, {Brickhouse}, {Liedahl}, \&
  {Raymond}}]{Smith01_APEC}
{Smith}, R.~K., {Brickhouse}, N.~S., {Liedahl}, D.~A., \& {Raymond}, J.~C.
  2001, \apjl, 556, L91, \dodoi{10.1086/322992}

\bibitem[{{Steiner} {et~al.}(2009){Steiner}, {Narayan}, {McClintock}, \&
  {Ebisawa}}]{Steiner09_simpl}
{Steiner}, J.~F., {Narayan}, R., {McClintock}, J.~E., \& {Ebisawa}, K. 2009,
  \pasp, 121, 1279, \dodoi{10.1086/648535}

\bibitem[{{Struble} \& {Rood}(1999)}]{zPers}
{Struble}, M.~F., \& {Rood}, H.~J. 1999, \apjs, 125, 35, \dodoi{10.1086/313274}

\bibitem[{{Taylor} {et~al.}(2006){Taylor}, {Gugliucci}, {Fabian}, {Sanders},
  {Gentile}, \& {Allen}}]{Perseus_Bcenter_Taylor06}
{Taylor}, G.~B., {Gugliucci}, N.~E., {Fabian}, A.~C., {et~al.} 2006, \mnras,
  368, 1500, \dodoi{10.1111/j.1365-2966.2006.10244.x}

\bibitem[{{Timmerman} {et~al.}(2021){Timmerman}, {van Weeren}, {McDonald},
  {Ignesti}, {McNamara}, {Hlavacek-Larrondo}, \&
  {R{\"o}ttgering}}]{minihalo-Pheonix}
{Timmerman}, R., {van Weeren}, R.~J., {McDonald}, M., {et~al.} 2021, \aap, 646,
  A38, \dodoi{10.1051/0004-6361/202039075}

\bibitem[{{Tribble}(1993)}]{Halo_Tribble93}
{Tribble}, P.~C. 1993, \mnras, 263, 31, \dodoi{10.1093/mnras/263.1.31}

\bibitem[{{T{\"u}mer} {et~al.}(2023){T{\"u}mer}, {Wik}, {Zhang}, {Hoang},
  {Gaspari}, {van Weeren}, {Rudnick}, {Stuardi}, {Mernier}, {Simionescu},
  {Rojas Bolivar}, {Kraft}, {Akamatsu}, \& {de Plaa}}]{Tumer23}
{T{\"u}mer}, A., {Wik}, D.~R., {Zhang}, X., {et~al.} 2023, \apj, 942, 79,
  \dodoi{10.3847/1538-4357/aca1b5}

\bibitem[{Virtanen {et~al.}(2020)Virtanen, Gommers, Oliphant, Haberland, Reddy,
  Cournapeau, Burovski, Peterson, Weckesser, Bright, {van der Walt}, Brett,
  Wilson, Millman, Mayorov, Nelson, Jones, Kern, Larson, Carey, Polat, Feng,
  Moore, {VanderPlas}, Laxalde, Perktold, Cimrman, Henriksen, Quintero, Harris,
  Archibald, Ribeiro, Pedregosa, {van Mulbregt}, \& {SciPy 1.0
  Contributors}}]{SciPy}
Virtanen, P., Gommers, R., Oliphant, T.~E., {et~al.} 2020, Nature Methods, 17,
  261, \dodoi{10.1038/s41592-019-0686-2}

\bibitem[{{Wik} {et~al.}(2009){Wik}, {Sarazin}, {Finoguenov}, {Matsushita},
  {Nakazawa}, \& {Clarke}}]{DanComa}
{Wik}, D.~R., {Sarazin}, C.~L., {Finoguenov}, A., {et~al.} 2009, \apj, 696,
  1700, \dodoi{10.1088/0004-637X/696/2/1700}

\bibitem[{{Wik} {et~al.}(2014){Wik}, {Hornstrup}, {Molendi}, {Madejski},
  {Harrison}, {Zoglauer}, {Grefenstette}, {Gastaldello}, {Madsen},
  {Westergaard}, {Ferreira}, {Kitaguchi}, {Pedersen}, {Boggs}, {Christensen},
  {Craig}, {Hailey}, {Stern}, \& {Zhang}}]{DanBullet}
{Wik}, D.~R., {Hornstrup}, A., {Molendi}, S., {et~al.} 2014, \apj, 792, 48,
  \dodoi{10.1088/0004-637X/792/1/48}

\end{thebibliography}
\bibliographystyle{aasjournal}

\newpage
\appendix
\section{Characterizing Crossarf Systematics} \label{CR systematics}

\begin{table}[t]
    \centering
    \caption{Observations Log For Crossarf Calculation Errors\label{tab:obs_cr}}
    \begin{tabular}{lcccc}
     \hline 
     \hline
     \\[-0.75em]
         Obs. ID & Date & RA & Dec & Exposure (ks) \\\\[-0.75em]
         \hline
         \\[-0.5em]
        10002016001 & 2012/07/08 & $166.11\degree$ & $38.21\degree$ & 44.06\\ 
        60002024002 & 2013/04/13 & $253.47\degree$ & $39.76\degree$ & 35.53\\ 
        60002047004 & 2013/10/05 & $38.2\degree$ & $20.29\degree$ & 38.24\\ 
        60002023022 & 2013/04/02 & $166.11\degree$ & $38.21\degree$ & 53.98\\ 
        60002024006 & 2013/07/12 & $253.47\degree$ & $39.76\degree$ & 20.6\\ 
        60002023012 & 2013/02/12 & $166.11\degree$ & $38.21\degree$ & 35.28\\ 
        60002024008 & 2013/07/13 & $253.47\degree$ & $39.76\degree$ & 20.61\\ 
        60002045004 & 2014/01/18 & $130.35\degree$ & $70.89\degree$ & 67.38\\ 
        60002047002 & 2013/10/02 & $38.2\degree$ & $20.29\degree$ & 32.43\\ 
        60160662002 & 2015/11/24 & $259.81\degree$ & $48.98\degree$ & 38.23\\ 
        90002003002 & 2015/01/24 & $110.47\degree$ & $71.34\degree$ & 32.18\\ 
        60002023010 & 2013/02/06 & $166.11\degree$ & $38.21\degree$ & 41.58\\ 
        60002020004 & 2013/12/31 & $194.05\degree$ & $-5.79\degree$ & 78.85\\ 
        60160272002 & 2015/12/06 & $98.94\degree$ & $-75.27\degree$ & 25.31\\ 
        60002020002 & 2013/12/16 & $194.05\degree$ & $-5.79\degree$ & 78.88\\ 
        60061027002 & 2013/02/14 & $40.27\degree$ & $-8.26\degree$ & 26.4\\ 
        60001101004 & 2015/01/18 & $36.27\degree$ & $18.78\degree$ & 70.56\\ 
        60001101002 & 2014/12/24 & $36.27\degree$ & $18.78\degree$ & 61.4\\ 
        60001042002 & 2013/02/06 & $68.3\degree$ & $5.35\degree$ & 38.04\\ 
        90202044002 & 2016/12/30 & $338.15\degree$ & $11.73\degree$ & 58.78\\ 
        10002015001 & 2012/07/07 & $166.11\degree$ & $38.21\degree$ & 78.45\\ 
        60101037004 & 2015/09/01 & $107.63\degree$ & $59.14\degree$ & 47.84\\ 
        60001099002 & 2013/12/17 & $327.98\degree$ & $-30.46\degree$ & 71.69\\ 
        60002024004 & 2013/05/08 & $253.47\degree$ & $39.76\degree$ & 55.0\\ 
        60160840002 & 2016/05/18 & $359.78\degree$ & $-30.63\degree$ & 38.17\\ 
      \hline
    \end{tabular}
\end{table}

\begin{figure}
\begin{center}
    \includegraphics[width=0.8\linewidth]{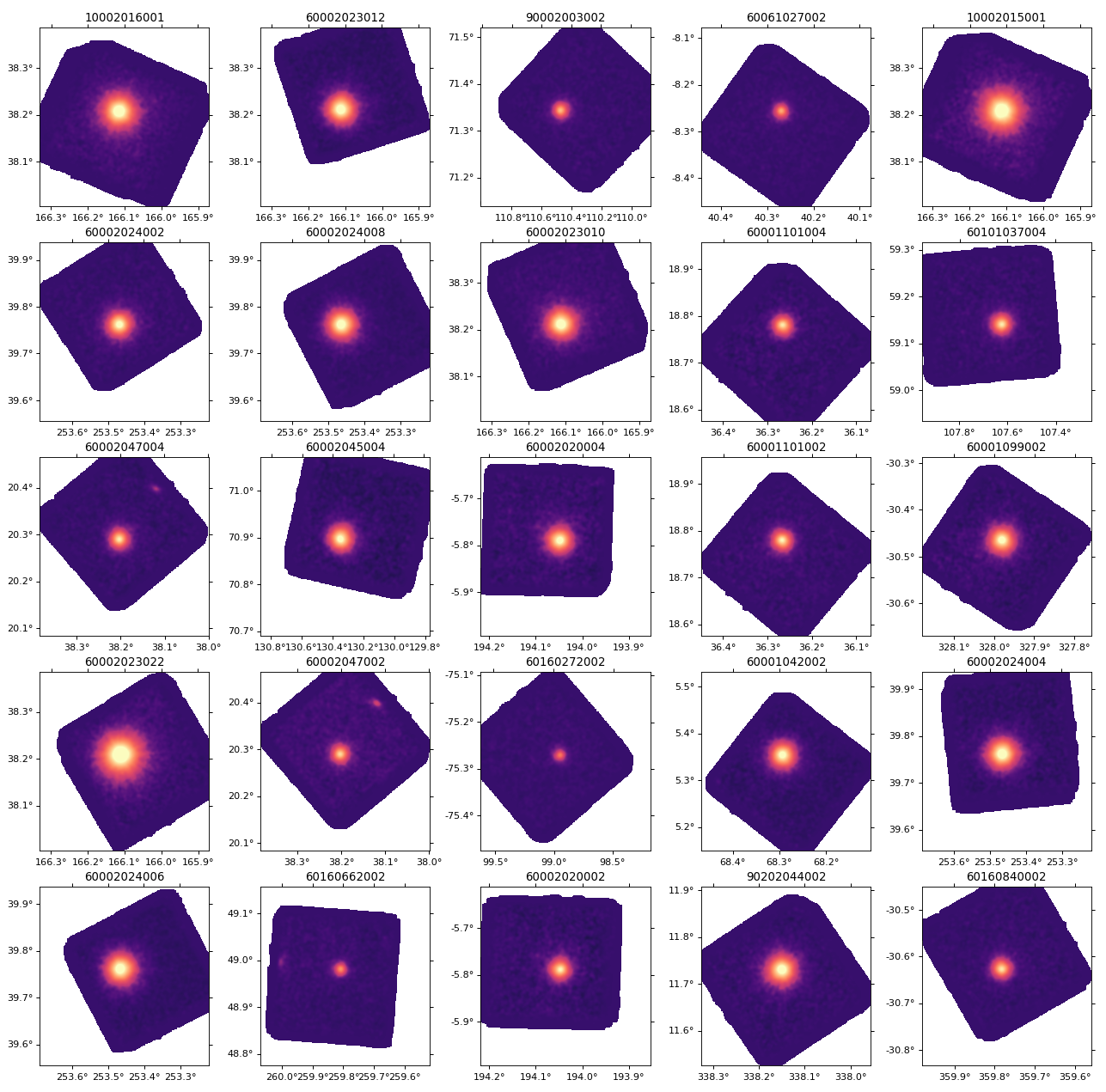}
\end{center}
    \caption{The observations used to estimate the accuracy of cross-ARF modeling. The observation IDs are listed along the top of the background-subtracted images; see table \ref{tab:obs_cr} for more information.}
    \label{fig:CR_obs}
\end{figure}

Due to \textit{NuSTAR's} large PSF (Half Power Diameter [HPD] of 58$^{\prime \prime}$ and full-width half maximum [FWHM] of 18$^{\prime \prime}$), photons from bright sources (such as the AGN at the center of the Perseus Cluster) can contaminate regions across the entire FOV. 
While \texttt{nuproducts} accounts for this scattering within a single region, it does not account for cross-talk between multiple sources of emission in separate regions. To correct for such cross-talk, this analysis makes use of the \texttt{nucrossarf} code.
\texttt{Nucrossarf} generates cross-ARF files that model the scattered emission from any number of source regions within the FOV into all defined spectral extraction regions. 
While this software has proven successful \citep[e.g.,][]{Tumer23}, its systematic uncertainty had not previously been characterized. In this section, we quantify that uncertainty.

To characterize \texttt{nucrossarf}'s ability to accurately model the scattered emission due to the wings of the PSF, we investigated the code's treatment of point sources. 
We selected 25 observations of beamed AGN (blazars) from the SWIFT/BAT catalog (see Table \ref{tab:obs_cr}), selecting for objects with Galactic latitude $|b| > 20 \degree$ to avoid dust-scattering halos that broaden the apparent PSF of Galactic sources.
For each point source, we fit the background using the methods of \citet{DanBullet}. 
Exposure maps were generated using \texttt{nuexpomap}, and spectra were extracted from both the background maps and the exposure-corrected data. 
The background-subtracted observations are shown in Figure \ref{fig:CR_obs}.

Once the data were reduced, we ran \texttt{nucrossarf} on two regions: a 30$^{\prime\prime}$ radius circle centered around the AGN (this is approximately equal to \textit{NuSTAR's} half power radius, so it should contain half the counts from a point source), and a concentric annulus from 30$^{\prime\prime}$ to 3$^\prime$. 
After obtaining ARF and RMF files for each region, we loaded the data from the inner region into {\tt XSpec} and fit the AGN with all model parameters freed.

In most cases, we fit the AGN to a powerlaw model.
An exception was the blazar Mrk 421. 
For observations of this well-studied object, we instead fit the spectra to a log-parabola model with a pivot energy fixed at $5$~keV, as described by \citet{MRK421}. 

Once the inner region's spectrum was fit, we loaded the spectrum and response files for the outer region. 
Ideally, the extrapolated model in the outer annulus should match the data without re-fitting. 
Discrepancies would stem from inaccurate cross-ARFs, which would be due to systematic uncertainties in the PSF modeling employed by \texttt{nucrossarf}.

\begin{figure}
\begin{center}
    \includegraphics[width=0.5\linewidth]{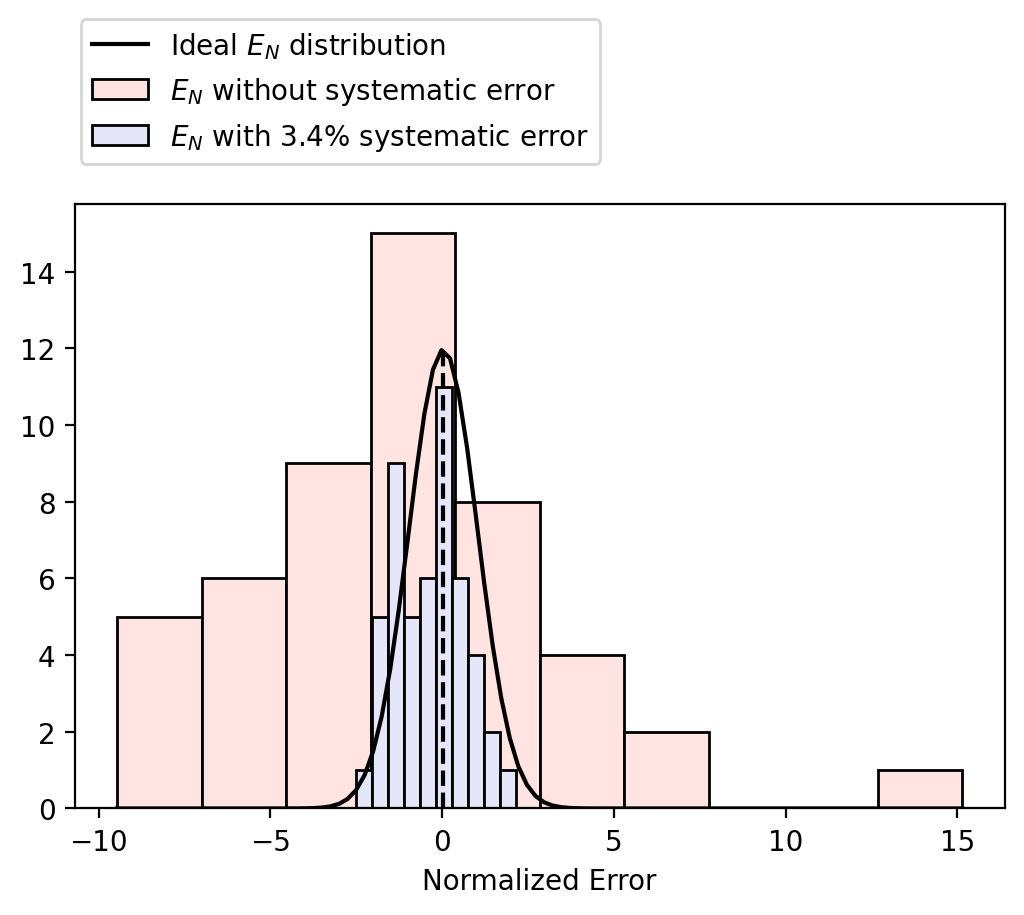}
\end{center}
    \caption{The normalized error $\text{E}_\text{N}$ (see Eq.~\ref{eq:err}) for 25 observations in both FPMA and FPMB (50 total) of point sources in the \textit{NuSTAR} archive. The solid black line shows an illustrative probability distribution function of a Gaussian with $\mu = 0$ (dashed black line) and $\sigma = 1$, which is the distribution of $E_N$ when all sources of error are accounted for. 
    The pink histogram is the distribution of $\text{E}_\text{N}$ when there is no systematic error accounted for. The purple histogram includes a 3.4\% systematic error, which produces the desired distribution of $\text{E}_\text{N}$. 
    The data appear to be negatively skewed, which would indicate that \texttt{nucrossarf} tends to over-model point source emission, but the skewdness (0.21) is not significant.}
    \label{fig:CR_err}
\end{figure}

We calculated the systematic uncertainty using the equation
    \begin{equation}\label{eq:err}
    \text{E}_\text{N} = \frac{(N_\text{data} - N_\text{model})}{\sqrt{ \sigma_\text{stat}^2 + \sigma_\text{sys}^2}}\, ,
    \end{equation}
where $\text{E}_\text{N}$ is the normalized error and $N_\text{model}$ and $N_\text{data}$ are the modeled and actual number of photons in the outer annulus, respectively.
If all sources of error are accounted for, the distribution of  $\text{E}_\text{N}$ 
for many observations should be a Gaussian with $\mu = 0$ and $\sigma = 1$. 
We considered two sources of error in our observations: statistical error ($\sigma_\text{stat} = \sqrt{N_\text{data}}$) and systematic error ($\sigma_\text{sys}$). 
We solved this equation on our data by finding a value of $\sigma_\text{sys}$ that produced a distribution of $\text{E}_\text{N}$ with $\mu = 0$ and $\sigma = 1$ (Figure \ref{fig:CR_err}).
To do this, we used the number of background-subtracted photons from the outer annulus ($N_\text{data}$) and the number of photons predicted by the model ($N_\text{model}$) and made an initial estimate for a value of $\sigma_\text{sys}$ to calculate $\text{E}_\text{N}$ for each source.
Once the distribution of $\text{E}_\text{N}$ was obtained, we fit it to a Gaussian model.
When the standard deviation of this distribution was greater/smaller than 1.0, we increased/decreased the value of $\sigma_\text{sys}$.
Following this treatment, we used the new $\sigma_\text{sys}$ to repeat the process until the standard deviation of $\text{E}_\text{N}$ was within $1.000 \pm 0.005$.
Once $\text{E}_\text{N} = 1.000 \pm 0.005$, we extracted the corresponding value of $\sigma_\text{sys}$.
Here, we report the systematic uncertainty as a relative standard error $\sigma$/N$_\text{data} \times 100$. 
However, depending on the methods used to fit our data, we obtained two different values of the systematic error. 
In method one, we froze all model parameters after fitting to the inner region, and we obtained $\sigma_\text{sys}$ based on the pure extrapolation into the outer region. This yielded a systematic uncertainty of $\sigma_\text{sys}/N \times 100 = 6.1\%$.

In method two, the normalization parameter was freed (and tied between the inner and outer regions) to vary while other parameters were fixed, and re-fit the model to our outer and inner regions simultaneously, yielding a systematic uncertainty of $\sigma_\text{sys}/N \times 100 = 3.4\%$. 
This method more accurately reflects the way that \texttt{nucrossarf} results are modeled, so we chose to use $3.4\%$ as the systematic error for the this analysis. This reflects the accuracy of \texttt{nucrossarf} to consistently model a point-like source between a typical circular region and a large outer annulus that captures emission from the wings of the PSF.

The $\text{E}_\text{N}$ distribution we obtained with $\sigma_\text{sys}/N \times 100 = 3.4\%$ is shown in Figure \ref{fig:CR_err}. The distribution does not have significant skewdness (0.21) or kurtosis (-0.34).

The exact level of uncertainty should be assessed on a case by case basis, as it may depend on the size and shape of spectral extraction regions and the nature of the source emission (point-like versus diffuse). Additionally, future work is required to characterize the uncertainties in different energy bands.

\end{document}